\documentstyle[12pt,fleqn,leqno]{article}
\begin{document}
\frenchspacing
\sloppy
\parskip=12pt
\baselineskip=20pt

$$
$$
\vspace{0.5cm}
{\begin{center}
{\LARGE{\bf{Gravity and Parity Violation}}} \\
\end{center}
\vspace{3.5cm}
\begin{center}
{{A. Geitner, D. Ketterer and H. Dehnen}} \\
\end{center}

\vspace{1cm}

\begin{center}
{\it{Fakult"t fr Physik\\
Universit"t Konstanz \\
Postfach 5560\\
D-78457 Konstanz}}
\end{center}
\newpage

\section*{Abstract}

Within a spin-gauge theory of gravity unified with the electroweak interaction we start with totally symmetric left- and right-handed
 fermions and explain the parity violation by symmetry breaking in
 such a way that the $W^{\pm}$-bosons couple only to the 
left-handed leptons. Right-handed neutrinos exist and couple as the right-handed electrons only
 to the Z-bosons (and to gravity). The mass of the neutrinos
 comes out to be necessarily zero. Therefore this procedure 
cannot be transferred to the quarks, because then the $u$-quark would become massless too; for this reason parity violation with
 respect to the quarks is avoided. On the other hand concerning
 gravity the $u$--quark couples only with  $u$--quarks and the $d$--quark only with $d$--quarks, however with the same strength, so
 that isotopic effects appear regarding the equivalence principle
 in such a way, that the macroscopic gravitational constant depends on the isotopic composition of the material. 
\newpage
\section*{1. Introduction}

In a previous paper Dehnen and Hitzer proposed in 1995 a
 unitary gauge theory of gravity based on the spin group
 SU(2)xU(1). The reason for doing this was the construction of a
 theory of gravitational interaction similar to the theories of
 the other fundamental physical interactions, i.e. the electroweak and strong
 interaction. In this way a natural unification of gravity with
 the other interactions should be possible and will be proposed in this
 paper. Also the problem of quantization of gravity may be solved
 in this way. Einstein's metric theory of gravity is considered as
 a pure classical theory with an effective non-euklidian metric
 deduced from the unitary gauge structure in a certain classical limit. 

The central idea for performing such a project is the fact, that gravity is marked out with respect to the elementary particles by
 non distinguishing between particles and antiparticles. Of
 course, there exist until now no direct experimental evidence for this; however one can show by gedanken-experiments that this must
 be valid; otherwise gravity would violate physical first principles (see e.g. Morrison, 1958; Nieto and Goldman, 1991;
 Dehnen and Ebner, 1996). Furthermore in the classical Einstein
 theory there exist also only geodesics and no antigeodesics. Then similar as in the weak interaction, where particles
indistinguishable with respect to the weak interaction, e.g. electron and neutrino, are put together in a (massless) isospin--dublett connected with the isospin group SU(2)xU(1), one has to put
 together with respect to gravity particles and antiparticles, e.g. electron and positron, in a spin--dublett connected with the
 spin group SU(2)xU(1). We call this a spin--dublett, because the
 combination of particles and antiparticles in a dublett is
 already realized in the chiral representation of Dirac's theory, where the 4-spinor is decomposed into two 2-spinors (Weyl spinors)
 representing the left- and right-handed particle/antiparticle
 dubletts. 

If one gauges the unitary groups in question the associated physical interactions result in form of bosonic gauge fields by
 introducing covariant derivations. However in the case of gauging
 the spin group one gets an additional interacting field, because  also the Dirac-matrices or the Weyl-matrices building  them up 
 become function valued and must be considered as Clifford algebra
 valued physical fields. Because these additional fields possess non trivial ground states, namely the standard representations of
 the Dirac- or Weyl-matrices, the Lagrangian for them must have the
 structure of a Higgs-field Lagrangian. It comes out, that the
 gravitational interaction is mediated exactly by these Clifford algebra valued Higgs-fields. In a certain classical limit
 Einstein's metric theory follows as an effective field theory, whereby Dehnen and Hitzer have restricted themselves to the linearized version. 

In the present paper we extend this procedure including the electroweak interaction of leptons and quarks. For this the Clifford
 algebra will be extended from the spin space to the spin-isospin
 space. We start with totally symmetric left- and right-handed
 fermionic multipletts. Simultaneously the parity violation of the weak interaction will be  generated by
 choosing the ground-states of the Clifford algebra valued Higgs--fields in such a way, that only the left--handed fermionic states
 couple to the $W^{\pm}$-bosons. Doing this generally one of the
 leptonic and hadronic particles remains massless. We identify
 this in the leptonic case with the left-- and right--handed
 neutrinos. In the hadronic case we renounce for this reason a
 parity violation because otherwise the $u$-quark would become massless too. 
 This procedure is not in contradiction with the experiments, because until
 now experimental evidence for parity violation exists only for the 
 weak interaction of the leptons. 

Concerning the gravitational interaction we have a situation
 similar to the previous paper of Dehnen and Hitzer (1995), namely that the Clifford algebra valued Higgs-fields
 mediate gravity. However we get a violation of the equivalence
 principle in such a way, that the gravitational interaction
 between leptons and hadrons is mediated by different Higgs-field
 components, so that the construction of a general non-euklidian
 metric did not succeed. A solution of this problem is an
 outstanding task. 

\section*{2. Spin-gauge Theory of Gravity}

At first we repeat briefly the SU(2)xU(1) spin-gauge theory of gravity so far as necessary. 

Form the beginning it is a Lorentz-invariant theory. The starting point is the Lagrange density of massless spin-1/2 particles 
$(\hbar = 1, c = 1)$
$$
{\cal{L}}_M = \frac{i}{2}\overline{\psi } \gamma ^\mu  \partial _\mu  \psi + 
\mbox{h.c.}
\leqno (2.1)
$$
where
$$
\gamma ^{(\mu } \gamma ^{\nu )} = \eta ^{\mu \nu } {\bf 1}
\leqno (2.1a)
$$
is valid  $(\eta ^{\mu \nu } = \eta _ {\mu \nu } =$ diag $(1,-1,-1,-1)$ Minkowski metric). Going over to the chiral representation of Dirac's $\gamma ^\mu $-matrices
$$
\gamma ^\mu  = \left( 
\begin{array}{ll}\displaystyle
0 & \sigma^\mu _L  \\
\sigma ^\mu _R & 0
\end{array}
\right) , \quad 
\psi = \left( 
\begin{array}{c}\displaystyle
\chi _R \\
\psi _L 
\end{array}
\right) 
\leqno (2.2)
$$
with the Weyl-matrices
$$
\sigma ^\mu _L = ({\bf 1} , - \sigma ^m ) , \quad \sigma ^\mu _R = ({\bf 1} , \sigma ^m )
\leqno (2.2a) 
$$
$(\sigma ^m , m \in  [1,2,3]$ Pauli-matrices)
we obtain from (2.1) and (2.1a): 
$$
{\cal{L}} _M = \frac{i}{2} (\chi _R ^{\dagger} \sigma ^\mu _R \partial _\mu \chi _R + 
\varphi ^{\dagger} _L \sigma ^\mu _L \partial _\mu  \varphi _L ) + h.c.,
\leqno (2.3)
$$
$$
\sigma ^{(\mu }_L \sigma ^{\nu )}_R = \eta ^{\mu \nu } {\bf 1} ,
\quad \sigma ^{(\mu }_R \sigma ^{\nu )}_L = \eta ^{\mu \nu } {\bf 1} \; .
\leqno (2.3a)
$$
The Weyl-spinor (2-spinor) $\chi _R$ represents the right-handed particle 
and left-handed antiparticle states and $\varphi _L$ the left-handed 
particle and right-handed antiparticle states, indistinguishable with 
respect to gravity. The unitary transformations taking into account this 
fact are the global SU(2)xU(1) 2-spinor transformations
$$
\begin{array}{l}
\chi ^\prime _R  =  U \chi _R , \quad \varphi ^\prime _L = U \varphi _L \\[0.4cm]
\sigma ^{\mu \prime }_R  =  U \sigma ^\mu _R U^{-1}, \quad \sigma ^{\mu \prime }_L = U \sigma ^\mu _L U^{-1} \\[0.4cm]
U  =  e^{i \lambda _j \tau ^j}, \quad \tau ^j = \frac{1}{2} \sigma ^j , \quad \sigma ^j = ({\bf 1}, \sigma ^i) \\[0.4cm]
j \in [0,1,2,3] , \quad i = \in [1,2,3],\;  \lambda _j = \mbox{const. (real valued)}
\end{array}
\leqno (2.4)
$$
($\sigma ^i$ Pauli-matrices), with 
respect to which (2.3) and (2.3a) are invariant or covariant respectively.

Now gauging this group by demanding $\lambda _j = \lambda _j (x^\mu )$ (real valued functions) the invariance of the Lagrange-density (2.3) is guaranteed furthermore by introducing the covariant derivative 
$$
\begin{array}{lll}
D_ \mu & = & \partial _\mu + i g \omega _\mu \;,  \\[0.4cm] 
\omega ^\prime  _\mu & = & U \omega _\mu U^{-1} + 
\frac{i}{g} U_ {|\mu }U^{-1}
\end{array}
\leqno (2.5)
$$
($g$ dimensionless gauge coupling constant) with the real-valued gauge-potentials $\omega _ {\mu j}$ defined by
$$
\omega _\mu = \omega _ {\mu j} \tau ^j . 
\leqno (2.5a)
$$
Simultaneously the Weyl matrices $\sigma ^\mu _L , \sigma ^\mu _R $ become function valued because of (2.4) and must be considered as additional fields denoted from now by $\, \tilde \sigma  ^\mu _L$ and $\, \tilde \sigma ^\mu _R$. For the Lagrangian of these we choose a Higgs-Lagrange density, because this fields possess a non-trivial ground-state, namely Weyl's standard representations (2.2a) with (2.3a). 

Thus the total Lagrange density consists of 3 terms: 
$$
{\cal{L}} = {\cal{L}}_M (\psi ) + {\cal{L}}_F (\omega ) + {\cal{L}}_H (\, \tilde \sigma )
\leqno (2.6)
$$
each gauge- and Lorentz-invariant:
$$
{\cal{L}}_M (\psi ) = \frac{i}{2}(\chi _R ^{\dagger} \, \tilde \sigma ^\mu _R D_ \mu \chi _R +
\varphi _L ^{\dagger} \, \tilde \sigma ^\mu _L D_ \mu \varphi _L ) + \mbox{h.c.},
\leqno (2.6a)
$$
$$
{\cal{L}}_F (\omega ) = - \frac{1}{16 \pi }F_ {\mu \nu j} F^{\mu \nu } _l \delta^{jl}
\leqno (2.6b)
$$
where
$$
F_ {\mu \nu j}\tau ^j = \frac{1}{ig} \left[ D_ \mu  , D_ \nu  \right] 
\leqno (2.6c)
$$
and 
$$
\begin{array}{l}
{\cal{L}}_H (\, \tilde \sigma ) = \mbox{tr}  \left\{ \left( D_ \alpha \, 
\tilde \sigma _ {\mu R}\right) \left( D^\alpha \, \tilde \sigma ^\mu_L \right) 
\right.  - \\[0.4cm]
- \left. \left( D_ \alpha \, \tilde \sigma _ {\mu R} \right) 
\left( D^\mu \, \tilde \sigma ^\alpha _L \right) - 
\left( D_ \alpha \, \tilde \sigma ^\alpha _ R \right) \left( D_ \beta 
\, \tilde \sigma ^\beta _L \right) \right\} -\\[0.4cm]
- \mu ^2 \mbox{tr} \left( \, \tilde \sigma ^\mu _L \, 
\tilde \sigma _ {\mu R} \right)  - 
\frac{\lambda }{12} \bigl[ \mbox{tr} (\, \tilde \sigma ^\mu _L 
\, \tilde \sigma _ {\mu R}) \bigr] ^2 - \\[0.4cm]
- k (\varphi ^{\dagger} _ L \, \tilde \sigma ^\mu _L 
\, \tilde \sigma _ {\mu R} \chi _R + \chi _R ^{\dagger} 
\, \tilde \sigma ^\mu _R \, \tilde \sigma _ {\mu L }
\varphi _L \left. \right) . 
\end{array}
\leqno (2.6d)
$$
In (2.6d) we have introduced also a Yukawa coupling term for producing the mass of the fermions after symmetry breaking $(k , \lambda > 0$ dimensionless, $\mu ^2 < 0$ with dimension of a mass square). 

The result of the theory described by (2.6) is the following  (Dehnen and Hitzer, 1995): The $\, \tilde \sigma $-Higgs-fields mediate a gravitational interaction between the fermions, which is in a classical limit in first order identical with Einstein's metrical theory, if we set for the ground state value
$$
- \frac{12 \mu ^2}{\lambda } = (2 \pi G)^{-1}
\leqno (2.7)
$$
($G$ Newtonian gravitational constant). Then the SU(2)-gauge bosons $\omega _ {\mu i}$ become massive with masses of the order of the Planck mass and can be therefore neglected in the low energy limit. However in the very early Universe they may be present and responsible for particle/antiparticle transitions explaining the inequilibrium of particles and antiparticles in the Universe. Only the U(1)-boson $\omega _ {\mu 0}$ remains massless and may be that of the hypercharge. This is the first hint  for unifying the spin-gauge theory of gravity with the isospin-gauge theory of the electroweak interaction. Because the spin-gauge theory of gravity is connected with the chiral representation of the fermionic states it may be possible to generate the parity violation of the weak interaction by such a unification of the gravitational and electroweak interaction choosing an appropriate symmetry breaking. 

On the other hand the gravitational theory presented above is a fermionic one from the very beginning and describes the gravitational interaction between elementary fermions only. However the gravitational action on bosons (e.g. light deflection) can be included subsequently, but will not be done in this paper. 
\section*{3. The Model}
In order to obtain a theory of gravity for leptons and quarks connected with the electroweak interaction we combine leptons and quarks into a quartett and restrict ourselves for simplicity to the first family:
$$
\psi _ {L, R} = 
\left( 
\begin{array}{c}
\nu \\
e\\
u\\
d
\end{array}
\right) _ {L, R} = 
\psi _ {L, R a} \; ,\quad a \in [1,2,3,4]
\leqno (3.1)
$$
Here $a$ is the isospinor index, which we omit if possible. Every element $\nu _ {L,R} = \nu _ {L,R \; A}, \quad e_ {L,R} = e_ {L,RA}, \; u_ {L,R} = u_ {L,RA}, \; d_ {L,R} = d_ {L,RA}$ is a 2-spinor in Dirac's chiral representation according to (2.2) \,  ($A \in [1,2]$ is the   2-spinor index and also omitted if possible). Of course, in a more sophisticated theory we must explain the ansatz (3.1) in such a way, that in consequence of a symmetry breaking the doublet $\nu \choose e$ does not underlie the strong interaction, i.e. the SU(3) colour group and $u \choose d$ it does; however this is not done in this paper, but may be possible. Restricting ourselves to the gravito-electroweak interaction we define the unitary transformations of (3.1) as follows
$$
\psi ^\prime  _ {L,R} = U \psi _ {L,R}, \quad U = e ^{i \lambda _j \tau ^j}
\leqno (3.2)
$$
with the generators $\tau ^{j}$ belonging to the group $SU(2)_{weak} \times U(1)_{hyperch.} \times SU(2)_ {spin}$:
$$
\tau ^j = \frac{1}{2} 
\left( 
\begin{array}{ll}
\sigma ^j & 0 \\
0 & \sigma ^j 
\end{array}
\right) 
{\bf 1}_ {spin}, 
\quad j = 1,2,3, 
\leqno (3.3a)
$$
$$
\tau ^{j = 0} = \tau ^H = \frac{1}{2}
\left( 
\begin{array}{ll}
{\bf 1}& 0 \\
0 & - \frac{1}{3}{\bf 1} \end{array}
\right) {\bf 1}_ {spin} ,
\leqno (3.3b)
$$
$$
\tau^j = \tau^{3+i} = \frac{1}{2}
\left( 
\begin{array}{ll}
{\bf 1} & 0 \\
0 & {\bf 1} 
\end{array}
\right) \sigma ^i _ {spin}, 
\quad i = 1,2,3 \; . 
\leqno (3.3c)
$$
The matrices in brackets act on the 4 isospin components of the isospinor 
(3.1), whereas the 2 $\times $ 2 spin-matrices act on the 2-spinors 
contained in (3.1). Furthermore the generators (3.3a) commute with those of (3.3b) and (3.3c) as well as (3.3b) with those of (3.3c). The isospin generators possess a block structure because of the broken symmetry between leptons and quarks. 

Of course, the generators (3.3) represent the restsymmetry group 
remaining after symmetry breaking of a larger group of the unified 
spin-isospin-space. Also this symmetry breaking is not performed in 
this paper, but we start with this restsymmetry. 

Gauging the group ($\lambda _j = \lambda_j (x^\mu ))$ we have to introduce 
the covariant derivative of the spin-isospinor (3.1): 
$$
D_ \mu \psi _ {L,R} = (\partial _\mu + i g_{(j)} \alpha _ {\mu j} \tau ^j) \psi _ {L,R}
\leqno (3.4)
$$
with 
$$
\alpha _ {\mu j} = W_ {\mu j}\; , \quad g_{(j)} = g_2, \quad j = 1,2,3 \; ,
\leqno (3.5a)
$$
$$
\alpha _ {\mu j } = B_ \mu \; , \quad g_ {(j)} = g_1, \quad j = 0 \; , 
\leqno (3.5b)
$$
$$
\alpha _ {\mu j} = \alpha _ {\mu 3 +i} = \omega _{\mu i}\; , \quad 
g_ {(j = 3 + i)} = g \;, \quad i = 1,2,3 \; .
\leqno (3.5c)
$$
Obviously, there is a total symmetry with respect to left- and right--handed states ``interacting'' with the weak $W$-bosons ($W_ {\mu 1}, W_ {\mu 2}, W_ {\mu 3}$; coupling constant $g_2$), the hypercharge-boson $B_ \mu $ (coupling constant $g_1$) and the spin-bosons $\omega _ {\mu i}$ (coupling constant $g$).

Now also the Weyl-matrices $\sigma ^\mu _ {L,R}$  must be considered as objects of the symmetry group (3.2), (3.3) in order to generate a gauge invariant Lagrangian. Doing this $\sigma ^\mu _ {L,R}$ will become also isospin valued for which we write
$$
\sigma ^\mu _ {L,R} = \sigma ^\mu _ {L, R\;A}{}^B \rightarrow  
\Sigma ^\mu _ {L,R \; A} {}^B {}_ a {} ^b ,
\leqno (3.6)
$$
where the capital indices are the spinorial ones, whereas the small latin indices belong to the isospin freedoms. The transformation law of (3.6) is that of a spin and isospin tensor and reads
$$
\Sigma ^\mu _ {R,L} {}^\prime = U \Sigma ^\mu _ {R,L} U^{\dagger}  , 
\leqno (3.7)
$$
in consequence of which $\Sigma ^\mu _ {R,L}$ will become 
function valued as before $\sigma ^\mu _ {L,R}$ in chapter 2. The field strength belonging to the gauge potentials 
$
\alpha _ {\mu j} = ( W_ {\mu j}, B_ \mu , \omega _ {\mu i}) 
$ are given by the usual definition
$$
\frac{1}{i } \left[D_ \mu , D_ \nu \right] = g_{(j)}
F_ {\mu \nu j}\tau ^j  .
\leqno (3.8)
$$

Now we are able to write down the total symmetric Lagrangian invariant with respect to the localized transformation group $SU(2)_ {weak}\times U(1)_ {hyperch.} \times SU(2)_ {spin}$ by generalization of (2.6):
$$
\begin{array}{l}
L = \frac{i}{2}\psi _L ^{\dagger} 
\Sigma ^\mu _L D_ \mu \psi _L + h.c. + \\[0.4cm] 
+ \frac{i}{2} \psi ^{\dagger} _ R 
\Sigma ^\mu _R D_ \mu \psi _R + h.c.
 + \\ [0.4cm] 
+ \mbox{tr}  \left\{  (D_ \alpha \Sigma ^\mu _R )(D^\alpha \Sigma _ {\mu L} ) - (D_ \alpha \Sigma ^\mu _ R )  (D_ \mu \Sigma ^\alpha  _L ) - 
(D_ \alpha \Sigma ^\alpha _ R ) (D_ \beta  \Sigma ^\beta  _ L ) \right\} - \\[0.4cm] 
- \mu ^2 \mbox{tr} \bigl( \Sigma ^\mu _ L \Sigma _ {\mu R}\bigr)  - \frac{\lambda }{12} \bigl[ \mbox{tr} (\Sigma ^\mu _L \Sigma _ {\mu R})\bigr]  ^2 - \\[0.4cm] 
- k \left[ \psi _L ^{\dagger}  \Sigma ^\mu _L \Sigma _ {\mu R} \psi _ R + 
\psi _R ^{\dagger}  \Sigma ^\mu _R \Sigma _ {\mu L} \psi _L \right] - \\[0.4cm] 
- \frac{1}{16 \pi } F_ {\mu \nu j } F^{\mu \nu } _
l \delta ^{jl} + L (\phi). 
\end{array}
\leqno (3.9)
$$
In contrast to (2.6) we have added a Lagrangian part $L(\phi)$ of an additional isospinorial Higgs-field $\phi$
$$
\begin{array}{l}
L(\phi) = \frac{1}{2}(D_ \mu \phi)^{\dagger}  D^\mu  \phi - 
\frac{\, \tilde \mu }{2}^2 \phi^{\dagger} \phi- 
\frac{\, \tilde \lambda }{4 !}(\phi^{\dagger} \phi)^2 \,   - \\[0.4cm] 
- \, \tilde k \left[(\psi _L ^{\dagger}  \phi) ( \phi^{\dagger}  \psi _R ) +
(\psi _R ^{\dagger}  \phi) (\phi^{\dagger}  \psi _L ) \right]. 
\end{array}
\leqno (3.9a)
$$
as in the standard-model of electroweak interaction. Otherwise we would get difficulties with the $\nu$-mass and that of the $Z$-boson; the $\nu$-mass would be imaginary and thus the neutrino unstable. The only nondimensionless constants are $\mu ^2 (< 0)$ and $\, \tilde \mu ^2 (< 0)$  (mass square) whereas $k, \, \tilde k, \lambda , \, \tilde \lambda $ are dimensionsless and positive. 

The covariant derivative of $\psi _ {L,R}$ in (3.9) is already given by (3.4); the covariant derivative of the tensorial quantity (with respect to spin- and isospin-space) $\Sigma ^\mu _ {L,R}$ reads
$$
D_ \alpha \Sigma ^\mu _ {L,R} = \partial _ \alpha  \Sigma ^\mu _ {L,R} + 
ig _{(j)} \alpha _{\alpha j} 
\left[\tau ^j , \Sigma ^\mu _ {L,R} \right], \quad j = 0, ... , 6.
\leqno (3.10)
$$
The Higgs-field $\phi$ in (3.9a) is considered as isospinor only $(\phi = \phi_a) $ with the transformation law
$$
\phi^\prime  = 
e^{i \lambda _j \tau ^j } \phi , \quad j = 0,1,2,3
\leqno (3.11a)
$$
and the covariant derivative
$$
D_ \mu \phi = \partial _\mu \phi+ 
i g_{(j)} \alpha _ {\mu  j} \tau ^j \phi, \quad j = 0,1,2,3.
\leqno (3.11b)
$$
Of course it would be necessary to consider $\phi$ also as a spinorial quantity, namely as a spinorial tensor in view of the Yukawa coupling term in (3.9a) (second line). But for simplicity we renounce  this  and keep in mind, that only the generators $\tau ^j$ with $j = 0,1,2,3$ act on $\phi$.

\section*{4. Symmetry Breaking}

In our model we have 2 Higgs-fields, namely $\Sigma ^\mu _ {L,R}$ and $\phi$. Therefore there exist 2 ground-state conditions for the minimum of energy given by the minimum of the Higgs-potentials  in (3.9) and (3.9a). These are
$$
\mu ^2 +
\frac{\lambda }{6} \mbox{tr}\bigl(  \stackrel{0}{\Sigma}{}^\mu _ L \stackrel{0}{\Sigma} _ {\mu R}\bigr)  = 0 
\quad (\mu  ^2 < 0 , \lambda > 0) \, , 
\leqno (4.1)
$$
and
$$
\, \tilde \mu ^2 + 
\frac{\, \tilde \lambda }{6} \stackrel{0}{\phi}{}^{\dagger} \stackrel{0}{\phi} = 0
\quad (\, \tilde \mu ^2 < 0 , \, \tilde \lambda > 0).
\leqno (4.2)
$$
Simultanously the field-equations are fulfilled herewith in absence of all fermions, gauge bosons and Higgs-particles (vacuum-state). In the first case (4.1) we can decompose the non-trivial ground states as follows:
$$
\stackrel{0}{\Sigma}{}^\mu _L = \stackrel{0}{\Sigma}{}^\mu {}_ {LA}{}^B {}_ a {}^b = 
\sigma ^\mu _ {LA} {}^B N_ {La}{}^b  \, , 
\leqno (4.3a)
$$
$$
\stackrel{0}{\Sigma}{}^\mu _R = 
\stackrel{0}{\Sigma}{}^\mu _ {RA} {}^B {}_ a {}^b
 = 
\sigma ^\mu _ {RA} {}^B N_ {Ra} {}^b \, . 
\leqno (4.3b)
$$
Herein the groundstates $\sigma ^{\mu }_ {L,R}$ are given by (2.2a) and $N_L$ and $N_ R$ must be chosen in such a way, that the right--handed leptons do not couple in (3.9) to the $W_ {\mu 1}, W_ {\mu 2}$-bosons (i.e. $W^{\pm}$), whereas the left--handed leptons couple; this is only the case for 
$$
N_L = N_ {La } {} ^b 
 = 
\left( 
\begin{array}{ll} 
l {\bf 1} & 0 \\
0 & q {\bf 1} 
\end{array}
\right) \, ,
\leqno (4.4a)
$$
$$
N_R = N_ {Ra} {}^b = 
\left( 
\begin{array}{ll}
-l \sigma ^3 & 0 \\
0 & q {\bf 1} 
\end{array}
\right) \,, 
\leqno (4.4b)
$$
where $l$ and $q$ are real valued constants connected with the lepton and quark masses respectively, see (5.8). Then  the parity violation is manifested in the commutator relations:
$$
\left[N_ L , \tau ^j \right] = 0, \quad \left[ N_ R, \tau ^j \right] \neq 0, \quad j = 1,2.
\leqno (4.4c)
$$
Of course by the choice of (4.4) the condition (4.1) can be fulfilled (see (4.5)). However beside (4.4) there exist other ground states (e.g. through substitution $\sigma ^3 \leftrightarrow {\bf 1}$), which cannot be transformed by a global unitary transformation into (4.4). Thus the theory possesses inequivalent vacuum states. 

As we will see later, the leptonic structure of (4.4) does not only generate parity violation but is also connected with vanishing $\nu $-mass. Therefore the leptonic structure cannot be transferred to the quark part in (4.4b); consequently the right--handed quarks will couple to the $W_ {\mu 1}$-, $W_ {\mu 2}$-bosons.

Insertion of (4.3) and (4.4) into (4.1) defines the ground state value $q$:
$$
\frac{8}{3}q^2 + 
\frac{\mu ^2}{\lambda } = 0 \, .
\leqno (4.5)
$$
If we indentify the excited Higgs-field of $\Sigma ^\mu _ {L,R}$ with the gravitational interaction as in chapter 2, $\sqrt {|\mu ^2 /\lambda |}$ will be of the order of the Planck mass (see (8.12)). Therefore the symmetry breaking producing parity violation happens at $10^{19}$ GeV, whereby the $\omega _ {\mu i}$-bosons become Planck-massive and play no role in the lower energy regions (see (6.9)). 

The second condition (4.2) will be fulfilled by the ansatz
$$
\stackrel{0}{\phi}= \stackrel{0}{\phi} _a = v N_a ,  \quad N_a = 
\left( \begin{array}{c}
1 \\
0\\
0\\
0
\end{array}
\right) 
\leqno (4.6)
$$
as in the standard model in view of the lepton masses. Insertion of (4.6) into (4.2) gives the second ground state value  $v$:
$$
\frac{1}{6} v^2 +
\frac{\, \tilde \mu ^2}{\, \tilde \lambda }= 0. 
\leqno (4.7)
$$
In this case $\sqrt {|\, \tilde \mu^2 /\, \tilde \lambda |}$ will be of the order of the $W^{\pm}$- and $Z^0$-boson masses. The final rest-symmetry is that of the electromagnetism.

Beyond the ground states we have to define the excited Higgs-field states. Starting with the $\Sigma ^\mu _ {L,R}$ fields it is to be taken into account, that they are hermitean with respect to the spinor and the isospinor components; thus we get 
$$
\Sigma^\mu _ {L,R A}{}^B {}_ a {}^b = 
\stackrel{0}{\Sigma}^\mu _ {L,RA} {}^B {}_ a {}^b + 
\varepsilon ^\mu _{L,R \nu r}(x^\alpha ) \sigma ^\nu _ {L,RA} {}^B N_ a ^{rb}
\leqno (4.8)
$$
where $N^r _a {}^b $ is a hermitean basis of the isospin-space and $\varepsilon ^\mu _ { L,R \nu r} (x^\alpha ) $ are the real valued excited fields; $\stackrel{0}{\Sigma} ^\mu _ {L,RA} {}^B _ a {}^b
$ are the ground-states (4.3). Already in the isospinorial ground-states $N_ {L,Ra}{}^b$ , see  (4.4), we must distinguish between lepton- and quark-part; therefore we have to do this also for $N^r _ a {}^b $ setting
$$
N^r_a {}^b = 
\left\{  
\begin{array}{lll}
N^l_a {}^b & = & 
\left( 
\begin{array}{ll}
\sigma ^l & 0 \\[0.4cm]
0 & 0 
\end{array}
\right), \quad \begin{array}{lll}
r & = & l \\
l & = & 0,1,2,3
\end{array}\\[0.4cm]
N^q_a {}^b & = & 
\left( 
\begin{array}{ll}
0 & 0 \\[0.4cm]
0 & \sigma^q 
\end{array}
\right) , \quad 
\begin{array}{lll}
r  & =  & 4 + q \\
q  & =  & 0,1,2,3 
\end{array}
\end{array}
\right.
\leqno (4.9)
$$
$(r = 0, ..., 7, \; \sigma ^0 = {\bf 1}) $, where $N^l_ a {}^b$ 
acts on leptons and $N^q_a {}^b$ on quarks only. 
Herewith the symmetry breaking between leptons and quarks is taken 
into account. 

The  separation between leptons and quarks is also the reason for the impossibility to reduce the Higgs-field $\phi$ by a unitary gauge to the ground-state (4.6). Thus we have for this:
$$
\phi_a = \stackrel{0}{\phi}_a + \varphi _a (x^\alpha ), 
\leqno (4.10)
$$
with the excited components $\varphi _a$.

\section*{5. The Field Equations for the Fermions}

The field equations for the fermions  following from the Lagrangian (3.9) take the form:
$$
\begin{array}{lll}
i \Sigma ^\mu _R D_ \mu \psi _R  & + & 
\frac{i}{2} (D_ \mu \Sigma ^\mu _R ) \psi _R - \\[0.4cm]
- k \Sigma ^\mu _R \Sigma _ {\mu L} \psi _L & - & 
\, \tilde k \phi( \phi^{\dagger}  \psi _L ) = 0
\end{array}
\leqno (5.1)
$$
and 
$$
\begin{array}{lll}
i \Sigma ^\mu _ L D_ \mu \psi _L & + & 
\frac{i}{2} (D_ \mu \Sigma ^\mu _L ) \psi _L -  \\[0.4cm]
- k \Sigma ^\mu _L \Sigma _ {\mu R}
\psi _R &  - & \, \tilde k \phi (\phi^{\dagger} \psi _R ) = 0
\end{array}
\leqno (5.2)
$$
as well as the adjoint equations.

In a first step we neglect the interaction with the excited Higgs-fields and use for $\Sigma ^\mu _ {L,R}$ and $\phi$ only the ground-states (4.3) and (4.6). Then it follows from (5.1) and (5.2) after insertion of the covariant derivatives (3.4), (3.5) and (3.10): 
$$
\begin{array}{l}
i \sigma ^\mu _R \partial _\mu N_R \psi _R - g_1 \sigma ^\mu _R B_ \mu \tau ^H N_R \psi _R - 
\\ [0.4cm]
- \frac{1}{2} g_2 \sigma ^\mu _R W_ {\mu j} \left\{  \tau ^j , N_R \right\} \psi _R - 
\\ [0.5cm]
- \frac{1}{2}g \omega _ {\mu j} \left\{  \tau ^j , \sigma ^\mu _R \right\} N_R \psi _R - 4 k N_R N_L \psi _L - \\ [0.5cm]
- \, \tilde k \phi(\phi^ {\dagger} \psi _L ) = 0
\end{array}
\leqno (5.3)
$$
and 
$$
\begin{array}{l}
i \sigma ^\mu _L \partial _\mu N_L \psi _L - g_1 \sigma ^\mu _L B_ \mu \tau ^H N_L \psi _L - 
\\ [0.5cm]
- \frac{1}{2} g_2 \sigma ^\mu _L W_ {\mu j} \left\{  \tau ^j , N_L \right\} \psi _L - 
\\ [0.5cm]
- \frac{1}{2}g \omega _ {\mu j} \left\{  \tau ^j , \sigma ^\mu _L \right\} N_L \psi _L - 4 k N_L N_R \psi _R - \\ [0.5cm]
- \, \tilde k \phi(\phi^ {\dagger} \psi _R ) = 0.
\end{array}
\leqno (5.4)
$$
The generators $\tau ^j $ and $\tau ^H$ are given by the definitions (3.3). Now we insert them together with the definitions (4.4) and (4.6); simultaneously we redefine the spinorial states (3.1) as follows: 
$$
\begin{array}{llllll}
\, \tilde \nu _ {L,R} & = & \sqrt l \nu _ {L,R}, \, & \tilde e_ {L,R} & = & \sqrt l  e_{L,R}  \, , \\[0.5cm]
\, \tilde u _ {L,R} & = & \sqrt q u _ {L,R} , & \, \tilde d _ {L,R} & = & \sqrt q d _ {L,R} \, , 
\end{array}
\leqno (5.5)
$$
in order to be able to compare them with the usual definition in the standard theory. 
In this way we obtain from (5.3) for the right--handed states: 
$$\begin{array}{l}
i \sigma ^\mu _R \partial _\mu \, \tilde \nu _R - 
\frac{1}{2}(g_1 B_ \mu + g_2 W_ {\mu 3}) \sigma ^\mu _R \, \tilde \nu _R - \\[0.4cm]
- \frac{1}{4}g \omega _ {\mu j} \left\{  \sigma ^j , \sigma ^\mu _R \right\} \, \tilde \nu _R - 
(4 kl - \, \tilde k \frac{v^2}{l} ) \, \tilde \nu _L = 0,
\end{array}
\leqno (5.6a)
$$

$$
\begin{array}{l}
i \sigma ^\mu _R \partial _\mu \, \tilde e_R - 
\frac{1}{2}(g_1 B_ \mu - g_2 W_ {\mu 3}) \sigma ^\mu _R \, \tilde e_R - \\[0.4cm]
- \frac{1}{4} g \omega _ {\mu j} \left\{  \sigma ^j , \sigma ^\mu _R \right\} 
\, \tilde e_R - 4 kl \, \tilde e _L = 0 , 
\end{array}
\leqno (5.6b)
$$

$$
\begin{array}{l}
i \sigma ^\mu _R \partial _ \mu \, \tilde u _R +
\frac{1}{2}( \frac{1}{3} g_1 B_ \mu - g_2 W_ {\mu 3}) \sigma ^\mu _R \, \tilde u _ R - \\[0.4cm]
- \frac{1}{2}g_2 (W_ {\mu 1} - i W_ {\mu 2}) \sigma ^\mu _R \, \tilde d _R - \\[0.4cm]
- \frac{1}{4}g \omega _ {\mu j} \left\{  \sigma ^j , \sigma ^\mu _R \right\} \, \tilde u _R - 4 k q \, \tilde u _L = 0 , 
\end{array}
\leqno (5.6c)
$$

$$
\begin{array}{l}
i \sigma ^\mu _R \partial _ \mu \, \tilde d _R + 
\frac{1}{2} (\frac{1}{3} g_1 B_ \mu + g_2 W_ {\mu 3}) 
\sigma ^\mu _ R \, \tilde d _R - \\[0.4cm]
- \frac{1}{2}g_2 (W_ {\mu 1} + i W_ {\mu 2} ) \sigma ^\mu _R \, \tilde u _R - \\[0.4cm]
- \frac{1}{4}g \omega _ {\mu j}\left\{  \sigma ^j , \sigma ^\mu _R \right\} \, \tilde d_R - 
4 k q \, \tilde d_L = 0 .
\end{array}
\leqno (5.6d)
$$
Obviously, the right--handed leptons do not couple to the $W_ {\mu 1}$- and $W_ {\mu 2}$-, i.e. the $W_ \mu ^{\pm}$-bosons describing the parity violation; however the right--handed quarks couple to $W^{\pm}$ with the same coupling constant as the left--handed quarks and leptons, see (5.7). The left--handed equations following from (5.4) are: 
$$
\begin{array}{l}
i \sigma ^\mu _L \partial _\mu \, \tilde \nu _L - 
\frac{1}{2}(g_1 B_ \mu + g_2 W_ {\mu 3}) \sigma ^\mu _L \, \tilde \nu _L - \\[0.4cm]
- \frac{1}{2}g_2 (W_ {\mu 1} - i W_ {\mu  2} )\sigma ^\mu _ L \, \tilde e _ L - \\[0.4cm]
- \frac{1}{4}g \omega _ {\mu j } \left\{  \sigma ^j  , \sigma ^\mu _L \right\} \, \tilde \nu _L - 
(\frac{\, \tilde k v^2}{l} - 4 kl ) \, \tilde \nu _R = 0, 
\end{array}
\leqno (5.7a)
$$

$$
\begin{array}{l}
i \sigma ^\mu _L \partial _ \mu \, \tilde e _L - \frac{1}{2}(g_1 B_ \mu - g_2 W_ {\mu 3}) \sigma ^\mu _L \, \tilde e _L - \\[0.4cm]
- \frac{1}{2} g_2 (W_ {\mu 1} + i W_ {\mu 2}) \sigma ^\mu _L \, \tilde \nu _L - \\[0.4cm]
- \frac{1}{4} g \omega _ {\mu j}\left\{  \sigma ^j , \sigma ^\mu _L \right\} \, \tilde e_L - 4 kl \, \tilde e_R = 0 ,\\[0.4cm]
\end{array}
\leqno (5.7b)
$$
$$
\begin{array}{l}
i \sigma ^\mu _L \partial _ \mu \, \tilde u _L + \frac{1}{2} (\frac{1}{3} g_1 B_ \mu  - g_2 W_ {\mu 3}) \sigma ^\mu _L \, \tilde u _L - \\[0.4cm]
- \frac{1}{2} g_2 (W_ {\mu 1} - i W_ {\mu 2}) \sigma ^\mu _L \, \tilde d_L - \\[0.4cm]

- \frac{1}{4} g \omega _ {\mu j} \left\{  \sigma ^j , \sigma ^\mu _L \right\} \, \tilde u _L - 4 k q \, \tilde u _ R = 0, 
\end{array}
\leqno (5.7c)
$$
$$
\begin{array}{l}
i \sigma ^\mu _L \partial _ \mu \, \tilde d_L + 
\frac{1}{2}(\frac{1}{3}g_1 B_ \mu + g_2 W_ {\mu 3}) \sigma ^\mu _L \, \tilde d _L -  \\[0.4cm]
- \frac{1}{2}g_2 (W_ {\mu 1} + i W_ {\mu 2}) \sigma ^\mu _ L \, \tilde u_L - \\[0.4cm]
- \frac{1}{4}g \omega _ {\mu j} \left\{   \sigma ^j , \sigma ^\mu _L \right\} \, \tilde d _ L - 4 kq \, \tilde d _R = 0. 
\end{array}
\leqno (5.7d)
$$
Evidently  for the masses of the fermions it follows:
$$
\begin{array}{lll}
m_ \nu & = & \frac{\, \tilde k v ^2}{l}- 4 kl = 4 kl - 
\frac{\, \tilde k v^2}{l} = 0, \\[0.4cm]
m_e & = & 4kl , \quad m_u = m_d = 4 kq .
\end{array}
\leqno (5.8)
$$
Accordingly the mass of the $u$- and $d$-quark is identical. This is approximately true with respect to the experiment, but it  can be changed by the choice of a more sophisticated ground state $N_ {L,R}$, see equation (4.4). The value  of $q$ is determined by (8.12) and is of the order of the Planck mass. Thus $k$ reduces the Planck mass to the quark mass according to (5.8) and $l$ to the electron mass. The neutrino mass vanishes identically as a consequence of the decoupling of the right--handed leptons from the $W^{\pm}$-bosons. 

In consequence of the vanishing neutrino mass 
$$
4 kl^2 - \, \tilde k v^2 = 0
\leqno (5.9)
$$
is valid, whereby the fermionic masses are connected with the boson  masses, see (6.8).  But there exist right- and left--handed neutrinos, from which only the left--handed one couples to the $W^{\pm}$-bosons; these are given by the usual combination: 
$$
W_ \mu ^{\pm} = W_ {\mu 1} \mp i W_ { \mu 2} \, . 
\leqno (5.10)
$$
The Weinberg mixture is the same as in the standard model, whereby the mass-matrix of the electroweak bosons will be diagonalized, see (6.8), 
$$
\begin{array}{lll}
W_ {\mu 3} & = & c_W Z_ \mu - s_W A_ \mu \, ,  \\[0.4cm]
B_ \mu & = & s_W Z_ \mu + c_W A_ \mu 
\end{array}
\leqno (5.11)
$$
where 
$$
\mbox{tg}  \theta _ W = \frac{g_1}{g_2}
\leqno (5.11a)
$$
($\theta _ W $ Weinberg angle, $s_W = \sin \theta  _W, c_W = \cos \theta _W$) and 
$$
g_1 c_W = g_2 s_W = e 
\leqno (5.11b)
$$
($e$ electric elementary charge). Thus we get in the equations (5.6) and (5.7) for the combinations of $B_ \mu $ and $W_ {\mu 3}$:
$$
\begin{array}{lll}
g_1 B_ \mu + g_2 W_ {\mu 3} & = & (g_1 s_W + g_2 c_W )Z_ \mu \, , \\[0.4cm]
g_1 B_ \mu - g_2 W_ {\mu 3} & = & (g_1 s_W - g_2 c_W )Z_ \mu + 2e A_ \mu \, , \\[0.4cm]
\frac{1}{3} g_1 B_ \mu - g_2 W_ {\mu 3} & = & ( \frac{1}{3}g_1 s_W - g_ 2 c_W )Z_ \mu + \frac{4}{3} e A_ \mu \, ,\\[0.4cm]
\frac{1}{3}g_1 B_ \mu + g_2 W_ {\mu 3} & = & (\frac{1}{3} g_1 s_W + g_2 c_W ) Z_ \mu - \frac{2}{3}e A_ \mu \, .
\end{array}
\leqno (5.12)
$$
Accordingly, left- and right--handed neutrinos couple in the same strenght to the $Z$-boson as also left- and right--handed electrons and quarks of the same sort. The electromagnetic interaction is the usual one. Additionally we have the interaction with the $\omega _ {\mu j}$-bosons in the same way for all leptons and quarks. However these bosons possess Planck masses, see (6.9) and (8.12), and can be therefore neglected in the low energy limit. However the coupling of the right--handed neutrino to the Z-boson may result in an experimental test of the theory.

\section*{6. The Field Equations for the Gauge Bosons}

The field equations for the gauge bosons following from the Lagrangian (3.9) take the form before symmetry breaking:
$$
\partial _\nu F^{\nu \mu i}  - g_ {(i)} f^{ij}{}_l \alpha_{ \nu j} F^{\nu \mu l} = 4 \pi  j^{\mu i}.
\leqno (6.1)
$$
Herein $f^{ij}{}_l$ are the structure constants belonging to the generators (3.3); they are
$$
f^{ij} {}_l = \left\{  \begin{array}{l}
\varepsilon ^{ij} {}_l \quad \left\{  \begin{array}{l}
i,j,l \quad \in  [1,2,3] \\[0.4cm]
i,j,l \quad \in  [4,5,6] 
\end{array}
\right. \\[0.4cm]
0 \quad \mbox{otherwise}
\end{array}
\right.
\leqno (6.1a)
$$
$
(\varepsilon^{ij} {}_l$ Levi-Civita-symbol). The currents consist of 3 gauge covariant parts belonging to the fermions and the $\Sigma ^\mu _ {L,R}$- and $\phi$-Higgs-fields:
$$
j^{\mu i} = j ^{\mu i} (\psi ) + j^{\mu i }(\Sigma ) +
j ^{\mu i}(\phi) .
\leqno (6.2)
$$
The fermionic part is given by
$$
j^{\mu i}(\psi ) = \frac{g_{(i)}}{2} \psi ^{\dagger} _L 
\left\{  \Sigma^\mu _L , \tau ^i \right\} \psi _L + 
\frac{g_ {(i)}}{2} \psi ^{\dagger} _ R \left\{  
\Sigma^\mu _R , \tau ^i \right\} \psi _R ,
\leqno (6.3)
$$
whereas the Higgs-field parts result after symmetry breaking into the mass term for the gauge bosons interacting with the excited Higgs-fields; restricting ourselves to the Higgs-field ground states we obtain:
$$
- 4 \pi (j^{\mu i}(\stackrel{\circ }{\Sigma} ) +
j^{\mu i} (\stackrel{ \circ }{\phi})) = M^{2 \mu }{}_\nu {}^{ij} \alpha ^\nu _j 
\leqno (6.4)
$$
with the mass-square matrix:
$$
\begin{array}{l}
M^{2 \mu }{}_\nu {}^{ij}  = 8 \pi g_ {(i)}g_ {(j)} \mbox{tr} \bigl\{  \bigl[
\tau ^{(i}, \stackrel{\circ }{\Sigma}^\mu _ L \bigr]\bigl[ \tau ^{j)}, \stackrel{\circ}{\Sigma}_ {\nu R}\bigr] \bigr.  +\\[0.4cm]
+ \bigl[ \tau ^{(i} , \stackrel{\circ}{\Sigma} ^\mu  _R \bigr]
\bigl[\tau  ^{j)}, \stackrel{\circ}{\Sigma }_ {\nu L} \bigr]- 
\delta ^\mu {}_\nu \bigl[ \tau ^{(i}, \stackrel{\circ}{\Sigma }^\alpha _ L
\bigr] \bigl[  \tau ^{j)}, \stackrel{\circ}{\Sigma}_ {\alpha R} \bigr] \bigl. \bigr\} 
 + \\[0.4cm]
+ 2 \pi g_ {(i)} g_ {(j)} \stackrel{\circ}{\phi} ^{\dagger} 
\bigl\{  \tau ^i , \tau^j \bigr\} \stackrel{\circ}{\phi} \delta ^\mu  {}_\nu 
\end{array} \, ,
\leqno (6.5)
$$
which is symmetric in $\mu , \nu $ and $i, j$. 
Here we remember that in the last term only the generators with $i,j \in [0,1,2,3]$ interact with $\stackrel{\circ}{\phi}$, see (3.11b). 

For $i,j \in [0,1,2,3]$, i.e. for the gauge-bosons $\alpha _ {\mu j} = \left\{  B_ \mu , W_ {\mu 1}, W_ {\mu 2}, W_ {\mu 3}\right\},$ see (3.3a,b) and (3.5a,b), the contribution of the $\stackrel{\circ}{\Sigma}^\mu _{L,R}$-Higgs-fields to the mass-square matrix vanishes; furthermore all components of the mass-square matrix are zero for the combination $i \in  [0,1,2,3]$, $j \in [4,5,6].$ Then we have only different from zero
$$
M^{2 \mu} {}_\nu {}^{ij} = 2 \pi g_ {(i)} g_ {(j)} \stackrel{\circ}{\phi}^{\dagger} \left\{  \tau ^i , \tau ^j \right\} \stackrel{\circ}{\phi} \delta ^\mu {}_ \nu 
\leqno (6.6)
$$
for $i,j \in [0,1,2,3]$ valid for the ``masses'' of $B_ \mu $ and $W_ {\mu 1}, W_ {\mu 2}$, $W_ {\mu 3}$ and 
$$
\begin{array}{l}
M^{2 \mu}  {}_ \nu {}^{ij} = 8 \pi g^2 \mbox{tr} \left\{  \left[\tau ^{(i}, \stackrel{\circ}{\Sigma }^\mu _ L \right.\right]  \left[ \tau ^{j)}, \stackrel{\circ}{\Sigma}  _{\nu R} \right] + 
\left[ \tau ^{(i}, \stackrel{\circ}{\Sigma}^\mu _R \right]\left[\tau^{j)}, \stackrel{\circ}{\Sigma}_ {\nu L} \right]- \\[0.4cm]
- \left. \delta ^\mu {}_ \nu \left[ \tau ^{(i}, \stackrel{\circ}{\Sigma}^\alpha _L \right] \left[ \tau ^{j)}, \stackrel{\circ}{\Sigma}_ {\alpha R} \right] \right\}  
\end{array}
\leqno (6.7)
$$
for $i,j \in  [4,5,6]$ valid for the masses of $\alpha _ {\mu j}= \alpha _ {\mu 3 + l} = \omega _ {\mu l} \quad (l = 1,2,3)$, see (3.3c) and (3.5c). 

With respect to the Higgs-field ground state $\stackrel{\circ}{\phi}$, see (4.6), the matrix (6.6) is identical with that in the electroweak standard model. Therefore the Weinberg mixing is the same as in the standard model, see (5.11), in order to diagonalize the matrix (6.6). In this way we have as usual:
$$
\left\{  
\begin{array}{l}
M_A  =  0 , \\[0.4cm]
M_ {W_1}  =  M_ {W_2} = M_ {W^{\pm}} = \sqrt \pi g_2 v , \\[0.4cm]
M_Z  =  v \sqrt{\pi (g^2 _1 + g_2 ^2 )} = M_ {W^{\pm}}/c_W . 
\end{array}
\right.
\leqno (6.8)
$$
On the other hand from (6.7) there follow the masses for the  $\omega _{\mu l}$-bosons mediating transitions between particles and antiparticles and violating the baryon/lepton number conservation. One finds in detail with the use of (2.2a), (4.3) and (4.4) taking the Pauli-matrices in their standard representation: 
$$
\begin{array}{l}
M^{2 \mu \nu k r} = 64 \pi g^2 q^2 \left\{  \eta ^{\mu \nu }\delta ^{kr } + \right.\\[0.4cm]
\left. + \delta ^{m n} \delta ^{kr} - \frac{1}{2} (\delta ^{m r} \delta  ^{nk} + \delta ^{mk} \delta ^{nr} ) \right\} \, , 
\end{array}
\leqno (6.9)
$$
where we have set $ i = 3 +k, \; j = 3 + r \quad (k,r  = 1,2,3).$ The fact, that here space-coordinates are marked out, depends on the chiral representation, where in the standard representation (2.2a) the $z$-axis is marked out also, as polarisation axis. Because $q$ will become later of the order of the Planck mass, see (8.12), the spin-gauge bosons $\omega_{\mu l}$ are super--heavy and can be neglected in the low energy limit. 

The field equations for the elektro--weak gauge bosons following from (6.1), (6.2) and (6.3) take the form with the use of (5.10) and (5.11):
$$
\begin{array}{l}
\partial ^\nu \partial _\nu A^\mu - \partial ^\mu \partial _\nu  A^\nu  + 
e \left\{  i \left[ (W^{\mu[ - } W^{+] \nu]}
 \right) _ {| \nu } \right. + \\[0.4cm]
\left. + W_ {\nu }^{[-} W^{+ ] \nu | \mu } - W^{[-}_\nu  W^{+ ]\mu |\nu  }
\right] - \\[0.4cm]
- e \left[W^{\mu (-} W_ \nu  ^{+)}A^\nu  + W^{\nu -} W_ \nu ^+ A^\mu 
\right] -\\[0.4cm]
- \left. 2 e \frac{c_W}{s_W} \left[ W^{\mu (-} W_ \nu ^{+)} Z^\nu 
 + W^{\nu -} W_ \nu  ^+ Z^\mu  \right] \right\} = \\[0.4cm]
= 4 \pi e \left\{  - \, \tilde e ^{\dagger} _L \sigma ^\mu _L 
\, \tilde e_L - \, \tilde e_R ^{\dagger} \sigma ^\mu _R \, \tilde e _R +
\frac{2}{3} \left( \, \tilde u_L ^{\dagger} \sigma ^\mu _L \, \tilde u_L + 
\right. \right. \\[0.4cm]
+\left.  \, \tilde u ^{\dagger} _R \sigma ^\mu _R \, \tilde u_R \right) - 
\frac{1}{3} \left( \, \tilde d_L ^{\dagger} \sigma ^\mu _L \, \tilde 
d _L +  \left. \, \tilde d_R ^{\dagger} \sigma ^\mu _R \, \tilde 
d_R \right) \right\} , 
\end{array}
\leqno (6.10)
$$
$$
\begin{array}{l}
\partial ^\nu \partial _\nu  Z^\mu - \partial ^\mu \partial \nu Z^\nu  + M^2 _Z Z^\mu - \\[0.4cm]
- 2 e \frac{c_W}{s_W} \bigl\{  i \bigl[ \bigl( W^{\mu [-} W^{+]\nu }\bigr) _ {|\nu } + W_ \nu ^{[-}W^{+] \nu |\mu } - \\[0.4cm]
- W_ \nu ^{[-} W^{+]\mu |\nu }  \bigr] - e \bigl[ W^{\mu (-} W_ \nu ^{+)} A^\nu +
W ^{\nu -} W_ \nu ^ + A^\mu  \bigr] - \\[0.4cm]
-2e \frac{c_W}{s_W} \bigl[
W^ {\mu (-} W_ \nu ^{+)} Z^\nu + W^{-\nu }  W_ \nu ^ + Z^\mu \bigr]  \bigr\} = \\[0.4cm]
= 4 \pi \bigl\{ \frac{1}{2} (g_1 {}s_W   +  g_2 c_W \bigr) \bigl[
\, \tilde \nu^{\dagger}  _L \sigma ^\mu _L \, \tilde \nu ^{\dagger} 
_L +
\, \tilde \nu _R \sigma ^\mu _R \, \tilde \nu _R \bigr] + \, \\[0.4cm]
+ \frac{1}{2}(g_1 s_W - g_2 c_W ) \bigl[
\, \tilde e _L \sigma ^\mu _L \, \tilde e _L + 
\, \tilde e _R \sigma ^\mu _R \, \tilde e _R \bigr]  - \\[0.4cm]
- \frac{1}{2} (\frac{1}{3}g_1 s_W - g_2 c_W) \bigl[\, \tilde u_L \sigma ^\mu _L \, \tilde u _L +
\, \tilde u _R \sigma ^\mu _R \, \tilde u _R \bigr] - \\[0.4cm]
- \frac{1}{2} (\frac{1}{3} g_1 s_W + g_2 c_W ) \bigl[ \, \tilde d _L \sigma ^\mu _L \, \tilde d_L +
\, \tilde d_R \sigma ^\mu _R \, \tilde d_R \bigr]  \bigr\} , 
\end{array} 
\leqno (6.11)
$$

$$
\begin{array}{l}
\partial ^\nu \partial _\nu W^{\mu +} - \partial^ \mu \partial _\nu W^{\nu+} 
+ M^2 _W W ^{\mu +} + \\[0.4cm]
+ 2 \bigl\{   - 2 e \frac{c_W}{s_W} i \bigl[  
\bigl( W^{+[\mu } Z^{\nu ]} \bigr) _ {|\nu } + \\[0.4cm]
+ W^{+[\mu |\nu ]}Z_ \nu - W_ \nu ^+ Z^{[\mu | \nu ]}  \bigr] - \\[0.4cm]
-ie \bigl[ \bigl( W^{+[\mu} A^{\nu ] } \bigr) _ {|\nu } + 
W^{+[\mu |\nu ]} A_ \nu - W_ \nu ^+ A^{[\mu |\nu ]} \bigr] +  \\[0.4cm]
+ 2 g_2 ^2 W_ \nu ^+ W^{+[ \mu } W^{\nu ]-} +
\bigl( 2e \frac{c_W }{s_W}\bigr) ^2 Z_ \nu Z^{[\nu } W^{\mu ]} + \\[0.4cm]
+ 2e^2 \frac{c_W }{s_W}\bigl( Z_ \nu A^{[\nu } W^{\mu ]+} +
A_ \nu Z^{[\nu } W^{\mu ]+} \bigr) + 
 e^2 A_ \nu A^{[\nu } W^{\mu ]+}  \bigr\} = \\[0.4cm]
= 
4 \pi g _2 \bigl\{   \, \tilde e_L ^+ \sigma ^\mu _L \, \tilde \nu _L
+  \bigl( \, \tilde d_L ^{\dagger} \sigma ^\mu _L \, \tilde u _L +
\, \tilde d _R ^{\dagger} \sigma ^\mu _R \, \tilde u_R \bigr)  \bigr\} ,
\end{array}
\leqno (6.12)
$$

$$
\begin{array}{l}
\partial ^\nu \partial _\nu W^{\mu -} - \partial ^\mu \partial _\nu  
W^{\nu -} + 
M _W ^2 W^{\mu -} + \\[0.4cm]
+ 2  \bigl\{  2 e \frac{c_W}{s_W} i \bigl[\bigl( 
W^{- [\mu }Z^{\nu ]}\bigr) _ {|\nu } + 
W ^{- [ \mu | \nu ]}Z_ \nu -\\[0.4cm]
-  W _ \nu  ^- Z^{[ \mu | \nu ]} \bigr] - i e \bigl[ \bigl( W^{- [ \mu } 
A^{\nu ]} \bigr) _ {|\nu } +  \\[0.4cm]
+ W ^{-[\mu |\nu ]}A_ \nu - W_ \nu ^- A^{[\mu |\nu ]}  \bigr]  + 
2 g^2 _2 W_ \nu ^- W^{+[\mu } W^{+ \nu ]+} + \\[0.4cm]
+ \bigl( 2 e  \frac{c_W}{s_W} \bigr) ^2 Z_ \nu Z^{[ \nu } W ^{\mu  ]-} +
2 e ^2 \frac{c_W}{s_W} \bigl(  Z_ \nu  A^{[\nu } W ^{\mu ]-} + \\[0.4cm]
+ A _ \nu Z^{[\nu } W ^{\mu ]-} \bigr) + 
e^2 A _\nu A ^{[ \nu }W^{\mu ]-} \bigr\} = \\[0.4cm]
= 4 \pi g_2 \bigl\{   \, \tilde \nu _L ^{\dagger} \sigma ^\mu _L \, 
\tilde e _L + 
 \bigl( \, \tilde u _L ^{\dagger} \sigma ^\mu _L \, \tilde d_L + \, 
\tilde u _R ^{\dagger} 
\sigma ^\mu _R \, \tilde d _R \bigr)  \bigr\}. 
\end{array}
\leqno (6.13)
$$
Herein the interaction with both Higgs-fields is neglected. 
The equations are more symmetric than in the standard model. 
We have right- and left--handed neutrinos; they couple in the neutral current of 
(6.11) with the same strength similarly as the left- and right--handed electrons;  
the coupling strength of electrons and neutrinos however is different. 
The same situation is valid for the quarks within the neutral current. The 
$W^\pm$-bosons however couple only to the left--handed leptons according to 
(6.12) and (6.13), which guarantees the parity violation. But the quarks 
couple to $W^\pm$ left- and right--handed and with the same strength as the left--handed leptons. As in the field equations of the fermions the coupling of the right--handed neutrinos to the Z-boson may represent a possibility of testing the theory.

Finally we give the field equations for the super-heavy spin-gauge bosons 
neglecting however nonlinearities, the interaction with the excited 
Higgs-fields included. From (6.1) it follows for $i,j \in [4,5,6]$ with the notation (3.5c) and with the use of (6.2) and (6.3): 
$$
\begin{array}{l}
\partial ^\nu \partial _\nu  \omega ^{\mu k } - \partial ^\mu 
 \partial _\nu \omega ^{\nu k} + M^{2 \mu \nu k r} \omega _ {\nu r} = \\[0.4cm]
= 4 \pi \frac{g}{2} \bigl[\psi _L ^{\dagger} \left\{  \sigma ^\mu _L , 
\tau ^k \right\} N_L \psi _L +  \\[0.4cm]
+ \psi _R ^ {\dagger} \left\{  \sigma ^\mu _R , \tau ^k \right\} N_R \psi _R 
 \bigr] . 
\end{array}
\leqno (6.14)
$$
The anticommutator within the currents takes the form: 
$$
\begin{array}{lll}
\left\{  \sigma ^\mu _L , \tau ^k \right\} & = 
& \left(   \sigma ^k , - \delta ^{m k } {\bf 1} \right) \\[0.4cm]
\left\{  \sigma ^\mu _R , \tau ^k \right\} & = & \left( 
\sigma ^k , \delta ^{m  k } {\bf 1} \right) . 
\end{array}
\quad \mu  = (0, \;  m \in \left[1,2,3 \right])
\leqno (6.14a)
$$
Obviously, transition currents between particle and antiparticles exist only for $k = 1,2$ and possess only the component $\mu  = 0$. On the other hand, as the $\mu  = 0$ component for $k = 3$ the 3-currents ($\mu = m \in [1,2,3])$ are all connected with particle-particle  and antiparticle-antiparticle transitions and have the structure $j^{mk} \sim \delta ^{mk}$; consequently also the adjoint gauge potentials are of the form $\omega ^{mk} \sim \delta ^{mk}$. This strong correlation between coordinates and group indices is a consequence of the chiral representation with the $z$-axis as spin polarisation axis. Herewith the mass term in (6.14) takes the form with respect to (6.9):
$$
M^{2 \mu \nu kr} \omega _ {\nu r} = 64 \pi g^2 q^2 (\omega ^{\mu k}+ \omega ^{mk}), 
\leqno (6.15)
$$
according to which the mass of the time--like component ($\mu  = 0$) of the gauge bosons is smaller than that of the space--like components $( \mu  = m \in [1,2,3]$) by the factor $1/\sqrt2$. However because $q \simeq m _ {Pl}$, see (8.12), the spin-gauge boson interaction can be neglected in the low energy limit; in the very early Universe it may have produced the particle-antiparticle inequilibrium in the Universe. 

\section*{7. The Field Equations for the Higgs-Fields}

In our theory there exist 2 different Higgs-fields $\phi$ and $\Sigma^\mu _ {L,R}$. 

a) The field equations for $\phi$ following from (3.9a) are:
$$
D_ \mu D^\mu \phi + \, \tilde \mu ^2 \phi +
\frac{\, \tilde \lambda }{6} (\phi^{\dagger} \phi) \phi = 
- 2 \, \tilde k \left[ (\psi _L ^{\dagger} \phi ) \psi _R  
+ (\psi _R ^{\dagger} \phi) \psi _L \right]
\leqno (7.1)
$$
as well as the adjoint equation. With the decomposition (4.10) and the 
ground state values (4.6) and (4.7) one finds for the excited Higgs-field 
$\varphi _a (x^\alpha )$ using (3.1) and neglecting the 
nonlinearities and the coupling to gauge-bosons:
$$
\partial ^\mu \partial _ \mu \varphi _a - 2 \, \tilde \mu^2 \varphi _1 \delta ^1 _ a = - 2 \, \tilde k v (\nu _L ^{\dagger} \psi _ {Ra} +
\nu ^{\dagger}_R \psi _ {La} ). 
\leqno (7.2)
$$
Accordingly only 
$\varphi _1 $ is massive and has the mass 
$$
M = \sqrt {-2 \, \tilde \mu ^2}
\leqno (7.2a)
$$ 
and is real valued. All other $\varphi _a \; (a = 2,3,4)$ are massless complex valued 
Goldstone--states. With respect to the right hand side of (7.2) the Goldstone--bosons $\varphi _a$ are connected with transitions from electrons and 
quarks into neutrinos (the adjoint $\varphi ^{\dagger a} $ are connected 
with the reverse processes). In view of (5.5), (5.8), (5.11b) and (6.8) the 
rigth hand side of (7.2) can be estimated as 
$$
\, \tilde k v (\nu _L ^{\dagger} \psi _ {Ra} + \nu _R ^{\dagger} 
\psi _ {La} ) 
\sim \frac{m_e}{M_ {W^\pm}} e( \, \tilde \nu _L
\, \tilde \psi _ {Ra} + \, \tilde \nu ^{\dagger} _R \, \tilde \psi _ {La} ) , 
\leqno (7.3)
$$
where $\, \tilde \nu $ and $\, \tilde \psi $ are the physical wave functions. 
Accordingly the coupling constant for the processes discussed above is very 
small, approximately smaller than the  electric one by the factor
$$
\frac{m_e}{m_ {W^\pm}} \simeq 5 \cdot 10^{-6} , 
\leqno (7.3a)
$$
so that these transitions may not be observable. 

b) In the second step we investigate the $\Sigma^\mu _ {L,R}$-Higgs-fields, which are connected with real gravitational interaction. The field equations following form (3.9) take the form 
$$
\begin{array}{l}
\partial _\alpha \partial ^\alpha \Sigma ^\mu _ {L,RAa}{}^B  {}^b -
2 \partial _\alpha \partial ^\mu \Sigma ^\alpha _ {L,RAa }
{}^B  {}^b + \\[0.4cm] 
+ \bigl(  \mu ^2 + 
\frac{\lambda }{6} \mbox{tr} \bigl(  \Sigma ^\alpha _L \Sigma _ {\alpha R} \bigr) \bigr) \Sigma _ {L,R}^\mu {} _A {}^B_a {}^b = \\ [0.4cm] 
= \frac{i}{2} \psi ^{\dagger} _ {R,L} {}^{Bb} \partial ^\mu \psi _ {LAa} + h.c. - \\[0.4cm] 
- k \bigl( \psi ^{\dagger} _ {L,R} {}^{Cd} \Sigma ^\mu _ {L,RC} {}^{Bb}_ d \psi _ {R,La} + 
\psi ^{\dagger Bb} _ {R,L} \Sigma ^{\mu } _ {L,RAa} {}^{Cd}\psi _ {L,RCd} \bigr) 
\end{array}
\leqno (7.4)
$$
where we have neglected the interaction with the gauge-bosons. Inserting the decomposition (4.8) and restricting ourselves for simplicity to the linear terms in $\varepsilon ^\mu _ {L,R \nu r} (x^\alpha )$ we obtain with the use of the groundstate conditions (4.1) and (4.3) 
$$
\begin{array}{l}
\partial _\alpha \partial ^\alpha \varepsilon ^\mu _ {L,R \nu r}
\sigma ^\nu _ {L,RA} {}^B N^r _a {}^b 
- 2 \partial _\alpha \partial ^\mu \varepsilon ^\alpha _ {L,R \nu r }\sigma ^ \nu _ {L,RA}{}^B N^r _ a {}^b + \\ [0.4cm] 
+ \frac{2}{3}\lambda  \bigl[ l \varepsilon _\alpha {}^\alpha {}_ {R_0} + q \varepsilon _ \alpha {}^\alpha {}_ {R4} - l \varepsilon _ \alpha {}^\alpha {}_ {L3} + q \varepsilon _ {\alpha } {}^\alpha {}_{L4} \bigr] 
 \sigma ^\mu _ {L,RA} {}^B N_ {L,Ra}{}^b = \\ [0.4cm] 
= \frac{i}{2} \psi ^{\dagger} _ {R,L} {}^{Bb}\partial _\mu  \psi _ {R,LA a} + \mbox{h.c.} \\[0.4cm] 
- k \bigl( \psi ^{\dagger} _ {L,R} {} ^{Cd} \sigma ^\mu _ {L,RC} {}^B N_ {L,Rd} {}^b \psi _ {R,LAa}  
+ \psi ^{\dagger} _ {R,L} {}^{Bb} \sigma ^\mu _ {L,RA} {}^C N_ {L,Ra} {}^d \psi _ {L,RCd} \bigr) . 
\end{array}
\leqno (7.5)
$$
Now, the spin- and isospin freedoms can be eliminated without loss of generality by tracing with the basis  elements (2.2a) and (4.9). In this way we get: 
$$
\begin{array}{l}
\partial _\alpha \partial ^\alpha \varepsilon ^\mu _ {L,R \nu r} - 
2 \partial _\alpha \partial ^\mu \varepsilon ^\alpha _ {L,R \nu r} + \\ [0.4cm] 
+ \frac{1}{3} \lambda \bigl[ l \varepsilon _ {\alpha } {}^\alpha {}_ {R_0} + q \varepsilon _\alpha {}^\alpha {}_ {R4} - 
l \varepsilon _ {\alpha } {}^\alpha {}_{L3}+ q \varepsilon _ {\alpha } {}^\alpha {}_{L4} \bigr] 
\bigl( N_ {L,Ra} {}^b N^{ra}_b \bigr)\delta _\nu {}^\mu   = \\[0.4cm] 
= \frac{i}{8} \psi ^{\dagger} _ {R,L} \sigma _ {\nu R, L}N^r \partial ^\mu \psi _ {R,L} + \mbox{h.c.} \\[0.4cm] 
- \frac{k}{4} \bigl[ \psi ^{\dagger} _ {L,R} {}^{Cd} \sigma ^\mu _ {L,RC} {}^B \sigma _ {\nu  R,LB} {}^A N_ {L,Rd} {}^b N^{ra} _b 
\psi _ {R,LAa} \\[0.4cm] 
+ \psi ^{{\dagger} B }_ {R,L}{}^b
 \sigma   _ {\nu R,LB} {}^A \sigma ^\mu _ {L,RA} {}^C N^r _ b {}^a N_ {L,Ra}{} ^d \psi _ {L, RCd} \bigr] . 
\end{array}
\leqno (7.6)
$$
For the $N$-matrix products in (7.6) it is valid according to (4.4) and (4.9) 
$$
\begin{array}{lll}
N_L N^r = N^r N_L & = & \left\{  
\begin{array}{l}
\left(   
\begin{array}{ll}
l \sigma ^l & 0 \\
0 & 0 
\end{array}
\right) \\ 
\left( 
\begin{array}{ll}
0 & 0 \\
0 & q \sigma ^q
\end{array}
\right) 
\end{array}
\right.
\\ \\[0.6cm] 
N_R N^r & = & \left\{  
\begin{array}{l}
\left( 
\begin{array}{ll}
-l \sigma ^3 \sigma ^l & 0 \\
0 & 0 
\end{array}
\right) \\
\left( 
\begin{array}{ll}
0 & 0 \\
0 & q \sigma ^{ q}
\end{array}
\right) 
\end{array}
\right. \\ \\[0.6cm] 
N^r N_R & = & \left\{  
\begin{array}{l}
\left( 
\begin{array}{ll}
-l \sigma ^l \sigma ^3 & 0 \\
0 & 0
\end{array}
\right) \\
\left( 
\begin{array}{ll}
0 & 0 \\
0 & q \sigma ^q 
\end{array}
\right) 
\end{array}
\right. 
\end{array}
\leqno (7.6a)
$$
with
$r  =  l , 
l  =  0,...3$, and $r  =  4 + q , 
q  =  0, ...3$. 
As one sees easily, the symmetric part of  (7.6) in $\mu , \nu $ has to do with the energy momentum tensor of the spinor fields and therefore with gravity, whereas the antisymmetric part is coupled with spin properties (see Dehnen and Hitzer, 1995). Consequently we restrict ourselves in the following to the symmetrized part of (7.6) taking the form:
$$
\begin{array}{l}\displaystyle
\partial _\alpha \partial ^\alpha \varepsilon ^{(\mu \nu )r} _ {L,R} - 2 \partial _\alpha \partial ^{(\mu }\varepsilon ^{\alpha \nu )r}_ {L,R} + \\ [0.4cm] 
+ \frac{1}{3}\lambda  \bigl[ l \varepsilon _\alpha {}^\alpha {}_ {R0} + q \varepsilon _\alpha {}^\alpha {}_ {R4}  - l \varepsilon _ {\alpha } {}^\alpha {}_{L3} + q \varepsilon _ {\alpha } {}^\alpha {}_{L4 } \bigr](N_ {L,Ra} {}^b N^{ra}_b )  \eta ^{\mu \nu } = \\ [0.4cm] 
= \frac{i}{8}\psi ^{\dagger} _ {R,L} \sigma ^{(\nu }_ {R,L} \partial ^{\mu )} N^r \psi _ {R,L} + h.c. \\[0.4cm] 
- \frac{k}{4} \bigl[\psi ^{\dagger} _ {L,R} N_ {L,R} N^r \psi _ {R,L} + \psi ^{\dagger} _ {R,L} N^r N_ {L,R}\psi _ {L,R} \bigr] \eta ^{\mu \nu } . 
\end{array}
\leqno (7.7)
$$
These equations decompose into 2 classes. For $r = 0,1,2,3$ 
we have energy and momentum of the leptons as sources for 
$\varepsilon ^{(\mu \nu )r}_ {L,R}$ and for $r = 4 + q$, $q = 0,1,2,3$ 
energy and momentum of the quarks. As one sees  from (4.9) and (7.6a) in the case 
of $l = 0,3$ and $q = 0,3$ there exist no transitions between 
leptons and quarks respectively, as it is the case in usual gravity; for 
$l = 1,2$ and $q = 1,2$ however a gravity like interaction exists with 
transition energy momentum tensors as source resulting in transitions 
between leptons and quarks respectively. But the transition probabilities are very small because the coupling constant for these processes is of the order of the gravitational constant; therefore we do not discuss them in the following.

\section*{8. Higgs-Field Gravity}

We investigate at first the ``quark-gravity'' 
$q = 0,3 \stackrel{\, \wedge }{=} r = 4,7$. 
From (7.7) it follows with the use of (7.6a):
$$
\begin{array}{l}\displaystyle
\partial _\alpha \partial ^\alpha \varepsilon ^{(\mu \nu ) }_ {L,R}{}^{4} - 2 \partial _\alpha \partial ^{(\mu } \varepsilon ^{\alpha \nu )}_ {L,R}{}^{4} - \\ [0.4cm] 
- \frac{\mu ^2}{4} \bigl[ \frac{l}{q} \bigl( \varepsilon _\alpha {}^\alpha {}_ {R} {}^0 - \varepsilon _ \alpha {}^\alpha {}_ {L} {}^3 \bigr) + 
\bigl( \varepsilon _\alpha  {}^\alpha_ {L } {}^{4}
+ \varepsilon _ \alpha  {}^\alpha _ {R}{}^{4}    \bigr) \bigr]\eta ^{\mu \nu } = \\[0.4cm]  
= \frac{1}{4q} \bigl\{  \frac{i}{2} \bigl[ \, \tilde u ^{\dagger} _ {R,L} \sigma ^{(\nu }_ {R,L} \partial ^{\mu )} \, \tilde u _ {R,L} + 
 \, \tilde d^{\dagger}  _ {R,L} \sigma ^{(\nu }_ {R,L} \partial ^{\mu )} \, \tilde d _ {R,L} \bigr] + \mbox{h.c.}- \\ [0.4cm] 
- \frac{m_ {u,d}}{4} \bigl[ \, \tilde u ^{\dagger} _ {L,R} \, \tilde u _ {R,L} + \, \tilde d^{\dagger}  _ {L,R} \, \tilde d _ {R,L} + \mbox{h.c.} \bigr]   \eta ^{\mu \nu } \bigr\}  
\end{array}
\leqno (8.1a)
$$
and 
$$
\begin{array}{l}
\partial _\alpha \partial ^\alpha \varepsilon ^{(\mu \nu )} _ {L,R}{}^7 - 2 \partial _ \alpha \partial ^{(\mu } \varepsilon ^{\alpha \nu)}_ {L,R} {}^7 = \\[0.4cm] 
= \frac{1}{4q} \left\{  \frac{i}{2} \bigl[
\, \tilde u ^{\dagger} _ {R,L}\sigma ^{(\nu }_ {R,L} \partial ^ {\mu )} 
\, \tilde u _ {R,L} - \right.  \, \tilde d ^{\dagger} _ {R,L} \sigma ^{(\nu }_ {R,L} \partial ^{\mu )} \, \tilde d _ {R,L} \bigr] + \mbox{h.c.}\\ [0.4cm] 
- \frac{m_ {u,d}}{4} \bigl[\, \tilde u ^{\dagger} _ {L,R} 
\, \tilde u _ {R,L} - \, \tilde d ^{\dagger} _ {L,R} \, \tilde d _ {R,L} + 
\mbox{h.c.} \bigr] \eta ^{m \nu }\left.  \right\} 
\end{array}
\leqno (8.1b)
$$   
where we have introduced the correct spinorial quantities according to (5.5) and $ \lambda q^2 = - \frac{3}{8}\mu ^2$ with respect to (4.5) as well as  $4 kq = m _ {u,d }$ in view of (5.8). On the right hand sides of (8.1)  one recognizes  the energy momentum tensor of the right/left--handed quarks
$$
T^{\mu \nu }_ {R,L}(u,d) = \frac{i}{2} \bigl( \, \tilde u ^{\dagger} _ {R,L} \sigma ^{( \nu }_ {R,L} \partial ^{\mu )} \, \tilde u_ {R,L} +\, \tilde d ^{\dagger} _ {R,L} \sigma ^{(\nu }_ {R,L} \partial ^{\mu )} \, \tilde d _ {R,L} \bigr) + \mbox{h.c.}; 
\leqno (8.2)
$$
its trace is using (5.6) and (5.7):
$$
T_ {R,L}(u,d) = \frac{m_ {u,d}}{2} \bigl( \, \tilde u ^{\dagger} _ R \, \tilde u _ L + 
\, \tilde d ^{\dagger} _ R \, \tilde d _L + h.c. \bigr). 
\leqno (8.2a)
$$
Furthermore because of $l/q = m_e /m_u \ll 1$ according to (5.8) we can neglect the first term in the bracket of the left hand side of (8.1a). Then we get for $ r = 4$: 
$$
\begin{array}{l}
\partial _\alpha \partial ^\alpha \varepsilon ^{(\mu \nu )}_ {L,R}{}^4 - 
2 \partial _\alpha \partial ^{(\mu } \varepsilon ^{\alpha \nu )} _ {L,R}{}^4 - \\[0.4cm] 
- \frac{\mu ^2}{4} \bigl[ \varepsilon _ {\alpha } {}^{\alpha }{}_L {}^4 + 
\varepsilon _ \alpha {}^\alpha {}_R {}^4 \bigr] \eta ^{\mu \nu } = \\[0.4cm] 
= \frac{1}{4q} \left\{  T^{\mu \nu }_ {R,L} (u,d) - \frac{1}{2} T_ {R,L} (u,d) \eta ^{\mu \nu } \right\} .
\end{array}
\leqno (8.3)
$$
This equation has the structure of Einstein's linearized field equations (c.f. also Dehnen and Hitzer, 1995 equ. (4.3), (4.4)); but in order to investigate this connection exactly it is necessary to analyze the backreaction of $\varepsilon ^{\mu \nu }_ {L,R} {}^{4,7}$ on the $u$- and $d$-quarks via the Weyl equations.

Starting from (5.1) and (5.2), neglecting the gauge-boson interaction and taking into account only linear terms of $\varepsilon ^{(\mu \nu )}_ {L,R} {}^{4,7}$ we obtain for the quarks 
$$
\begin{array}{l}
i \sigma ^\mu _ {R,L} \partial _ \mu {{\, \tilde u} \choose {\, \tilde d}} _ {R,L} + 
\frac{i}{q}\bigl( \varepsilon ^{(\mu \nu ) \; 4}_ {R,L} \pm
\varepsilon ^{(\mu \nu ) \; 7}_ {R,L} \bigr) \sigma _ {\nu  R,L} \partial _ \mu  {{\, \tilde u} \choose {\, \tilde d}} _ {R,L}  + \\ [0.4cm] 
+ \frac{i}{2} \partial _ \mu \frac{1}{q}\bigl( 
\varepsilon ^{(\mu \nu ) \, 4 } _ {R,L} \pm \varepsilon ^{(\mu \nu ) \, 7}_ {R,L} \bigr) \sigma _ {\nu R,L} {{\, \tilde u} \choose {\, \tilde d}}_{R,L} \\ [0.4cm] 
- m_ {u \atop d} \bigl[ 1 + \frac{1}{4q} \bigl( 
\varepsilon ^{\alpha }_ {L,R \alpha } {}^4 \pm \varepsilon ^\alpha _ {L,R \alpha }{}^7 \bigr) + \\ [0.4cm] 
+ \frac{1}{4q}  \bigl( \varepsilon ^\alpha _ {R,L \alpha } {}^4 \pm \varepsilon ^\alpha _ {R,L \alpha } {}^7 \bigr) \bigr]  
{{\, \tilde u} \choose {\, \tilde d}} _ {L,R} = 0 \\[0.4cm] 
\end{array}
\leqno (8.4)
$$
where (4.3), (4.4) (4.9), (5.5) and (5.8) have been used. Evidently on the $u$-quark acts only the combination
$$
\varepsilon ^{(\mu \nu )} _ {R,L} {}^{+} = \frac{1}{q} \bigl( \varepsilon ^{(\mu \nu )}_ {R,L}{}^4 + \varepsilon ^{(\mu \nu)}_ {R,L}{}^7 \bigr)  
\leqno (8.5a)
$$
and on the $d$-quark 
$$
\varepsilon ^{(\mu \nu )\;- }_ {R,L} = \frac{1}{q} \bigl( 
\varepsilon ^{(\mu \nu ) \; 4} _ {R,L} - \varepsilon ^{(\mu \nu) }_ {R,L} {}^7 \bigr) .
\leqno (8.5b)
$$
Herewith (8.4) takes the form
$$
\begin{array}{l}
i \sigma ^\mu _ {R,L} \partial _ \mu {{\, \tilde u} 
\choose {\, \tilde d}} _ {R,L} +
 i \varepsilon ^{(\mu \nu )}_ {R,L} {}^\pm
\sigma  _ {\nu R,L} \partial _ \mu 
{{\, \tilde u} \choose {\, \tilde d}} _ {R,L} + \\[0.4cm] 
+ \frac{i}{2} \partial _ \mu \varepsilon ^{(\mu \nu )} _ {R,L} {}^\pm
 \sigma _ {\nu R, L} {{\, \tilde u} \choose {\, \tilde d}} _ {R,L}  - 
 \\[0.4cm] 
- m_ {u \atop d} \bigl[ 1 + \frac{1}{4} \bigl( 
\varepsilon  _ {\alpha}{}^\alpha {}_{L,R} 
 {}^\pm + \varepsilon  _ {\alpha}{}^\alpha {}_{R, L} {}^\pm \bigr) \bigr] 
  {{\, \tilde u} \choose {\, \tilde d}} _ {L,R} 
= 0
\end{array}
\leqno (8.6)
$$
On the other hand the field equations for 
$\varepsilon ^{(\mu \nu )}_ {R,L} {}^\pm$ follow by addition and 
substraction of 
the equations (8.3) and (8.1b)  for $r = 4,7$. In this way we obtain:
$$
\begin{array}{l}
\partial _\alpha \partial ^\alpha
\varepsilon ^{(\mu \nu )}_ {R,L} {}^{+} - 2 \partial _ \alpha 
\partial ^{( \mu } \varepsilon ^{\alpha \nu )}_ {R,L} {}^{+} -  \\[0.4cm] 
- \frac{\mu ^2 }{4q} \bigl( \varepsilon _ {\alpha}{}^\alpha {}_{L}{}^4  +
\varepsilon  _ {\alpha}{}^\alpha {}_{R}{}^4
 \bigr) \eta  ^{\mu \nu}  = \\[0.4cm] 
= \frac{1}{2 q ^2} \left\{  T^{\mu \nu }_ {L,R} 
(u) - \frac{1}{2} T_ {L,R} (u) \eta ^{\mu \nu } \right\} , 
\end{array}
\leqno (8.7a)
$$

$$
\begin{array}{l}
\partial _ \alpha \partial ^\alpha \varepsilon ^{(\mu \nu )}_ {R,L} {}^- - 2 \partial _ \alpha \partial ^{(\mu } \varepsilon ^{\alpha \nu)}_ {R,L} {}^- - \\[0.4cm] 
- \frac{\mu ^2 }{4q} \bigl( \varepsilon  _ {\alpha}{}^\alpha {}_{L} {}^4 + \varepsilon  _ {\alpha}{}^\alpha {}_{R}{}^4  \bigr) \eta ^{\mu \nu } = \\[0.4cm] 
= \frac{1}{2 q^2} \left\{  T^{\mu \nu }_ {L,R} (d) - \frac{1}{2}
T_ {L,R} (d) \eta ^{\mu \nu } \right\} 
\end {array}
\leqno (8.7b)
$$
with the definition (8.2). Obviously according to (8.6) and (8.7) the 
$u$-quarks interact only with $u$-quarks and $d$-quarks with $d$-quarks, however with the same strength. On the other hand it is a gravitational interaction coupling to the energy momentum-tensor but without universality on the microscopic level. Such difficulties with the equivalence--principle appear also in other attempts of quantum gravity (see e.g. Scherk, 1981). Only if the matter consists of equal parts of $u$- and $d$-quarks, i.e. neutrons and protons,  universality is guaranteed exactly; then $\varepsilon ^{(\mu \nu )}_ {L,R} {}^7$ vanishes (see (8.1b)) and the universal interaction is mediated by
$
\varepsilon ^{(\mu \nu )}_ {L,R} {}^{+} = \varepsilon ^{(\mu \nu) }_ {L,R} {}^- = \frac{1}{q} \varepsilon ^{(\mu \nu )}_ {L,R} {}^4
$ with the field equation (8.3).

In view of the macroscopic limit we restrict ourselves to this last case and 
consider left- and right-handed states as equally represented, i.e.
$$
\begin{array}{l}
T^{\mu \nu }_R (u,d) = 
T ^{\mu \nu }_L (u,d) = 
\frac{1}{2} T^{\mu \nu } (u,d)
\\[0.4cm]
\frac{1}{q} \varepsilon ^{(\mu \nu )}_L {}^4 = \frac{1}{q} \varepsilon ^{(\mu \nu )}_R {}^4 = \varepsilon ^{(\mu \nu )} . 
\end{array}
\leqno (8.8)
$$
 Then by addition of the left- and right-handed equation (8.3) one obtains 
$$
\begin{array}{l}
\partial _\alpha \partial ^\alpha \varepsilon ^{(\mu \nu )}  
- 2 \partial _ \alpha \partial ^{(\mu }\varepsilon ^{\alpha \nu )} - \\[0.4cm] 
- \frac{\mu ^2}{2} \varepsilon _ \alpha {}^\alpha \eta ^{\mu \nu } = 
\frac{1}{8 q^2} \left\{  T^{\mu \nu} (u,d) - 
\frac{1}{2} T(u,d) \eta ^{\mu \nu } \right\} . 
\end{array}
\leqno (8.9)
$$
Simultaneously, the Weyl matrices can be combined to generalized Dirac-matrices
$$
\, \tilde \gamma ^\mu = \bigl( \eta ^{\mu \nu } + 
\varepsilon ^{(\mu \nu )} \bigr) 
\left(
\begin{array}{ll}
0 & \sigma_ {\nu L} \\
\sigma _ {\nu R} & 0 
\end{array}
\right) 
\leqno (8.10)
$$
so that the Weyl-equations (8.6) for the quarks interacting with $\varepsilon ^{(\mu \nu )} $ take the form of generalized Dirac equations:
$$
\begin{array}{l}
i \bigl[ \, \tilde \gamma ^\mu  \partial _ \mu 
{{\, \tilde u} \choose {\, \tilde d}} + 
\frac{1}{2} \partial _ \mu \, \tilde \gamma ^\mu 
{{\, \tilde u} \choose {\, \tilde d}} \bigr] - \\ [0.4cm] 
- \frac{1}{4} m_ {u,d} \, \tilde \gamma ^\alpha \, \tilde \gamma _ \alpha  {{\, \tilde u} \choose {\, \tilde d}} = 0. 
\end{array}
\leqno (8.11)
$$
The field equations (8.9) and (8.11) are identical with those in the paper of Dehnen and Hitzer, 1995 (equ. (5.5) and (5.7)). There is shown, that herewith Einstein's linearized gravitational theory is achieved with the metric and the gravitational constant: 
$$
\left\{  \, \tilde \gamma ^\mu , \, \tilde \gamma ^\nu \right\} = 2 g^{\mu \nu } {\bf 1} , \quad \frac{1}{q^2} = 
64 \pi G 
\leqno (8.12)
$$
($G$ Newtonian gravitational constant). 

In a second step we investigate the ``lepton-gravity'' $l = r = 0,3$. From (7.7) it follows with the use of (7.6a):
$$
\begin{array}{l}
\partial _ \alpha \partial ^\alpha \varepsilon _ L ^{(\mu \nu) 0 }
- 2 \partial _ \alpha \partial ^{(\mu }\varepsilon _L ^{\alpha \nu ) 0} - \\ [0.4cm] 
- \frac{1}{4} \mu ^2 \frac{l}{q} \bigl[\frac{l}{q}  (\varepsilon _ {\alpha} {}^\alpha {}_R {}^0 - 
\varepsilon _ \alpha {}^\alpha {}_L {}^3 \bigr) + 
\bigl( \varepsilon _\alpha {}^\alpha {}_R {}^4 + 
\varepsilon _ \alpha {}^\alpha {}_L {}^4 \bigr) \bigr] \eta ^{\mu \nu } = \\[0.4cm] 
= \frac{1}{4l} \left\{  \frac{i}{2} \bigl[ \, \tilde \nu _R ^{\dagger} 
\sigma ^{(\nu }_R \partial ^{\mu )} \, \tilde \nu _R + 
\, \tilde e _ R ^{\dagger} \sigma ^{(\nu } _R 
\partial ^{\mu )} \, \tilde e_R \bigr]  + h.c. \right. - \\[0.4cm] 
 - \frac{m_e }{4} \bigl[
\, \tilde \nu _L ^{\dagger} \, \tilde \nu _R + 
\, \tilde e _L ^{\dagger} \, \tilde e_R + h.c. \bigr] \left. \eta ^{\mu \nu }
 \right\} , 
\end{array}
\leqno (8.13a)
$$

$$
\begin{array}{l}
\partial _\alpha \partial ^\alpha \varepsilon ^{(\mu \nu )0} _R - 
2 \partial _ \alpha \partial ^{(\mu}   \varepsilon _R ^{\alpha \nu ) 0} = \\[0.4cm] 
= \frac{1}{4l} \bigl\{  \frac{i}{2} \bigl[ \, \tilde \nu ^{\dagger} _L \sigma ^{(\nu } _L \partial ^{\mu)} \, \tilde \nu _L + 
\, \tilde e_L ^{\dagger} \sigma ^{(\nu }_L \partial ^{\mu )} \, \tilde e_L \bigr]  + 
h.c.  -  \\[0.4cm] 
- \frac{m_e}{4} \bigl[- \, \tilde \nu _R ^{\dagger} \, \tilde 
\nu _L + \, \tilde e _R ^{\dagger} \, \tilde e_L  + h. c. \bigr] 
\eta ^{\mu \nu } \bigr\} , 
\end{array}
\leqno (8.13b)
$$

$$
\begin{array}{l}
\partial _\alpha \partial ^\alpha \varepsilon ^{(\mu \nu)3 }_L -
2 \partial _ \alpha \partial ^{(\mu }\varepsilon ^{\alpha \nu ) 3}_L = \\[0.4cm] 
= \frac{1}{4l} \bigl\{  \frac{i}{2} \bigl[ \, \tilde \nu _ R ^{\dagger}  \sigma _R ^{(\nu } \partial ^{\mu )} \, \tilde \nu _R - 
\, \tilde e _R ^{\dagger} \sigma ^{(\nu }_R  \partial ^{\mu )} \, \tilde e_R \bigr] + \mbox{h.c.}  \\[0.4cm] 
- \frac{m_e}{4} \bigl[ \, \tilde \nu _L ^{\dagger} \, \tilde \nu _R 
- \, \tilde e_L ^{\dagger} \, \tilde e_R + \mbox{h.c.} \bigr]
 \eta ^{\mu \nu }\bigr\} , \\
\end{array}
\leqno (8.13c)
$$
$$\begin{array}{l}
\partial _\alpha \partial ^\alpha \varepsilon ^{(\mu \nu )3}_R - 
2 \partial _\alpha \partial ^{(\mu } \varepsilon ^{\alpha \nu )3}_R +  \\[0.4cm] 
+ \frac{1}{4} \mu ^2 \frac{l}{q} \bigl[\frac{l}{q}\bigl( \varepsilon _\alpha {}^\alpha  {}_R {}^0 - \varepsilon _ \alpha {}^\alpha {}_L {}^3 \bigr) + \bigl( \varepsilon _a {}^\alpha {}_R {}^4 + 
\varepsilon _ \alpha {}^\alpha {}_L {}^4 \bigr) \bigr] \eta ^{\mu \nu } = \\[0.4cm] 
= \frac{1}{4l} \bigl\{ \frac{i}{2} \bigl[ \, \tilde \nu _L ^{\dagger}  \sigma _L ^{(\nu }\partial ^{\mu )}\, \tilde \nu _L - 
\, \tilde e _L ^{\dagger} \sigma _ L ^{(\nu } \partial ^{\mu )} \, \tilde e_L \bigr] + \mbox{h.c.}  \\[0.4cm] 
+ \frac{m_e}{4} \bigl[
\, \tilde \nu _R ^{\dagger} \, \tilde \nu _L + \, \tilde e _R ^{\dagger} 
\, \tilde e _L + \mbox{h.c.} \bigr] \eta ^{\mu \nu } \bigr\} 
\end{array}
\leqno (8.13d)
$$
where we have used (4.5), (5.5) and (5.8). 

Now we investigate the action of $\varepsilon ^{\mu \nu } _ {L,R} {}^{0,3}$ on the leptons. From (5.1) and (5.2) we obtain neglecting the gauge--boson interaction and taking into account only linear terms of $\varepsilon ^{\mu \nu }_ {L,R} {}^{0,3}$ : 
$$
\begin{array}{l}
i \sigma ^{\mu }_L \partial _ \mu {{\, \tilde \nu }
 \choose {\, \tilde e}} _ L + 
\frac{i}{l} \bigl( \varepsilon_L  ^{(\mu \nu )0} \pm 
\varepsilon_L  ^{(\mu \nu )3} \bigr) \sigma _ {\nu L}  
{{\, \tilde \nu } \choose {\, \tilde e}} _ L - \\[0.4cm] 
+ \frac{i}{2l}\partial _\mu \bigl( \epsilon ^{(\mu \nu )0}_L \pm \epsilon ^{(\mu \nu )3}_L \bigr) \sigma _ {\nu L}  {{\, \tilde \nu } \choose {\, \tilde e}} _ L 
- m_e \bigl[ {0 \choose {\, \tilde e}_R } +  \\[0.4cm] 
+ \frac{1}{4l} \bigl( \varepsilon _ \alpha {}^\alpha {}_R {}^0 \pm 
\varepsilon _ \alpha {}^\alpha {}_R {}^3 \mp \varepsilon _ \alpha{}^\alpha  {}_L 
{}^0- 
\varepsilon _ \alpha {}^\alpha {}_L {}^3 \bigr)  
 {{\, \tilde \nu } \choose {\, \tilde e}} _ R \bigr] = 0 
\end{array}
\leqno (8.14a) 
$$
and
$$
\begin{array}{l}
i \sigma ^\mu _R \partial _ \mu 
{{- \, \tilde \nu } \choose {\, \tilde e}} _ R + \frac{i}{l}
\bigl( \varepsilon ^{(\mu \nu )0}_R \pm \varepsilon ^{(\mu \nu )3}_R \bigr) \sigma _ {\nu R} 
{{ \, \tilde \nu } \choose {\, \tilde e}} _ R +  \\[0.4cm] 
+ \frac{i}{2l} \partial _ \mu \bigl( \varepsilon ^{(\mu \nu )0}_R \pm 
\varepsilon ^{(\mu \nu )3}_R \bigr) \sigma _ {\nu R} 
{{ \, \tilde \nu } \choose {\, \tilde e}} _ R - \\[0.4cm] 
- m_e \bigl[ {{ 0 } \choose {\, \tilde e}} _ L + 
\frac{1}{4e} \bigl( \varepsilon _ {R \alpha } ^\alpha {}^0
 \pm \varepsilon _ {R \alpha }^\alpha {}^3 - \varepsilon _ {L \alpha }^\alpha {}^3 \mp \\[0.4cm] 
\mp \varepsilon _ {L \alpha } ^\alpha {}^0 \bigr) {{ \, \tilde \nu } \choose {\, \tilde e}} _ L \bigr] = 0 . 
\end{array}
\leqno (8.14b)
$$
Evidently, on the neutrinos acts only the combination
$$
 \, \tilde \varepsilon ^{(\mu \nu) }_ {R,L} {}^{+} = 
\frac{1}{l} \bigl( \varepsilon ^{\mu \nu }_ {R,L}{}^0
+ \varepsilon ^{(\mu \nu )}_{R,L} {}^3 \bigr) 
\leqno (8.15a)
$$
and on the electrons
$$
\, \tilde \varepsilon ^{(\mu \nu )}_ {R,L}{}^- = 
\frac{1}{l} \bigl( 
\varepsilon ^{\mu \nu }_ {R,L}{}^0 - 
\varepsilon ^{(\mu \nu )}_{R,L} {}^3 \bigr) . 
\leqno (8.15b)
$$
Obviously the neutrinos interact only with neutrinos and electrons only with electrons according to (8.13) and (8.14). It may be of interst that such violations of the equivalence principle, where different leptons underlie different gravitational field strengths, have been suggested by Butler et al. and Pantaleone et al. in 1993 and by Bahcall et al. in 1995 in order to generate neutrino oscillations with massless neutrinos for solving the solar neutrino problem. 

Restricting ourselves finally to the ``electron-gravity'' we get from (8.13) and (8.14) 
$$
\begin{array}{l}
i \sigma ^{\mu }_ {L,R} \partial _ \mu \, \tilde e _ {L,R} + 
\frac{i}{2} \, \tilde \varepsilon ^{(\mu \nu )}_ {L,R} {}^- 
\sigma _ {\nu L,R} \, \tilde e _ {L,R} + \\[0.4cm] 
+ \frac{i}{2} \partial _ \mu \, \tilde \varepsilon ^{(\mu \nu )}_ {L,R } {}^- \sigma _ {\nu L,R} \, \tilde e _ {L,R} - \\[0.4cm] 
- m_e \bigl[1 + \frac{1}{4} \bigl( \, \tilde \varepsilon  _ {\alpha}{}^\alpha {}_{R}{}^- + \, \tilde \varepsilon  _ {\alpha}{}^\alpha {}_{L}{}^- \bigr) \bigr] \, \tilde e_ {R,L} = 0
\end{array}
\leqno (8.16)
$$
and 
$$
\begin{array}{l}
\partial _ \alpha \partial ^\alpha \, \tilde \varepsilon ^{(\mu \nu )}_ {L,R}{}^- - 2 \partial _ \alpha \partial ^{(\mu } 
\, \tilde \varepsilon  ^{\alpha \nu )}_ {L,R}{}^-  - \\[0.4cm] 
- \frac{1}{4}  \frac{\mu ^2}{q} \bigl[ \frac{l}{q} (\varepsilon _ \alpha {}^\alpha {}_R  {}^0 - \varepsilon _ {\alpha }{}^\alpha {}_L {}^3 \bigr) + 
\varepsilon _ {\alpha }{}^\alpha {}_R {}^4 + 
\varepsilon _\alpha  {}^\alpha {}_L {}^4 \bigr] \eta ^{\mu \nu } = \\[0.4cm] 
 = \frac{1}{2 l^2} \bigl\{ T^{\mu \nu }_ {R,L} (\, \tilde e ) - 
 \frac{1}{2}T_ {R,L}(\, \tilde e ) \eta ^{\mu \nu } \bigr\} 
\end{array}
\leqno (8.17)
$$
where
$$
\begin{array}{lll}\displaystyle
T^{\mu \nu }_ {L,R} (\, \tilde e) & = & \frac{i}{2}\, \tilde e_ {L,R}^{\dagger} 
\sigma ^{(\nu }_ {L,R} \partial ^{\mu )} \, \tilde e _ {L,R} + \mbox{h.c.}, \\[0.4cm] 
T_ {L,R} (\, \tilde e ) & = & \frac{m_e}{2} \, \tilde e_R ^{\dagger} \, \tilde e_L + \mbox{h.c.}
\end{array}
\leqno (8.18)
$$
is the energy momentum tensor of the electrons and its trace. These equations correspond exactly to those for the quarks given by (8.6) and (8.7) and describe a gravitational interaction between the electrons on the microscopic level. 

The comparison from (8.7) and (8.17), c.f also (8.9), shows that the macroscopic gravitational constant for the electrons is by the factor $\frac{q^2}{l^2} = \bigl( \frac{m_u}{m_e} \bigr) ^2$ larger than that for the quarks, see (8.12). This means that the strength of the gravitational attraction between two electrons is exactly the same as between two quarks. This leads together with the
 gravitational $u$--$u$ and $d$--$d$ attraction to isotopic effects
 with respect to the equivalence principle in such a way that the
 effective macroscopic gravitational constant depends on the
 isotopic composition of the material; this would result into a
 ``fifth force'' as suggested by Fischbach in 1986. However, because up today no experimental evidence for this exists (see Adelberger, 1991), our theory is in the present form not in agreement with the experiments concerning the equivalence principle. We hope that this lack can be avoided by a suitable modification of the theory. 

On the other hand  there exists a not understood phenomenon. The different measurements of the macroscopic gravitational constant
 show relative differences up to $7,4 \times 10 ^{-3}$, although
 the relative accuracy of the measurements is of the order of
 $10^{-4}$ (see e.g. Gillies, 1997). This means that large unknown systematic errors are
 involved. In this connection we want to point to the fact, that
 relative differences of the measured gravitational constant of the mentioned magnitude would be  compatible with our isotopic effect. However this seems to be in contradiction to the E"tv"s experiments. 

\section*{References}
\begin{itemize}
\item[{}] Adelberger, E. G. et al. (1991). {\sl{Ann. Rev. Nucl. Part. Sci.}}, {\bf{41}}, 269.
\item[{}] Bahcall, J. N. et al. (1995). {\sl{Phys. Rev. D}}, {\bf{52}}, 1770.
\item[{}] Butler, M. N. et al. (1993). {\sl{Phys. Rev. D}}, {\bf{47}}, 2615.
\item[{}] Dehnen, H., and Hitzer, E. (1995). {\sl{Int. J. theor. Phys.}}, {\bf{34}}, 1981.
\item[{}] Dehnen, H. and Ebner, D. (1996). {\sl{Foundation of Physics}}, {\bf{26}}, 105. 
\item[{}] Gillies, G. T. (1997). {\sl{Rep. Prog. Phys.}}, {\bf{60}}, 151.
\item[{}] Morrison, P. (1958). {\sl{American Journal of Physics}}, {\bf{26}}, 358.
\item[{}] Nieto, M., and Goldman, T. (1991). {\sl{Physics Report}}, {\bf{205}}, 221.
\item[{}] Fischbach, E. et al. (1986). {\sl{Phys. Rev. Lett.}}, {\bf{56}}, 3.
\item[{}] Pantaleone, J. et al. (1993). {\sl{Phys. Rev. D}}, {\bf{47}}, 4199.

\item[{}] Scherk, J. (1981). {\sl{Unification of the fundamental particle interactions, eds. S. Ferrar, J. Ellis and P. Nieuwenhuizen}}, Plenum New York, p. 381. 
See also {\sl{Phys.  Lett}}, {\bf{88B}}, 265 (1979).
\end{itemize}
\end{document}